\documentclass[useAMS,usenatbib]{mn2e}
\usepackage{graphicx}
\usepackage{rotating}

\title[Constraints on large scale inhomogeneities from WMAP-5 and
SDSS]{Constraints on large scale inhomogeneities from WMAP-5 and SDSS:
  confrontation with recent observations} \author[Paul Hunt and Subir
Sarkar]{Paul Hunt$^{1}$\thanks{E-mail:
    Paul.Hunt@fuw.edu.pl; s.sarkar@physics.ox.ac.uk} and Subir Sarkar$^{2}$\\
  $^{1}$Institute of Theoretical Physics, Warsaw
  University, ul Ho\.za 69, 00-681 Warsaw, POLAND\\
  $^{2}$Rudolf Peierls Centre for Theoretical Physics, University of
  Oxford, 1 Keble Road, Oxford OX1 3NP, UK}
\begin{document}



\maketitle

\label{firstpage}

\begin{abstract}
  Measurements of the SNe~Ia Hubble diagram which suggest that the
  universe is accelerating due to the effect of dark energy may be
  biased because we are located in a 200-300 Mpc underdense `void'
  which is expanding 20-30\% faster than the average rate. With the
  smaller global Hubble parameter, the WMAP-5 data on cosmic microwave
  background anisotropies can be fitted without requiring dark energy
  if there is some excess power in the spectrum of primordial
  perturbations on 100~Mpc scales. The SDSS data on galaxy clustering
  can also be fitted if there is a small component of hot dark matter
  in the form of 0.5 eV mass neutrinos. We show however that if the
  primordial fluctuations are gaussian, the expected variance of the
  Hubble parameter and the matter density are far too small to allow
  such a large local void. Nevertheless many such large voids have
  been identified in the SDSS LRG survey in a search for the late-ISW
  effect due to dark energy. The observed CMB temperature decrements
  imply that they are nearly empty, thus these real voids too are in
  gross conflict with the concordance $\Lambda$CDM model. The recently
  observed high peculiar velocity flow presents another challenge for
  the model. Therefore whether a large local void actually exists must
  be tested through observations and cannot be dismissed {\em a
    priori}.
  \end{abstract}

\begin{keywords}
  cosmic microwave background, cosmological parameters, cosmology:
  theory, dark matter, large-scale structure of Universe
\end{keywords}

\section{Introduction}

The Einstein-de Sitter (E-deS) universe with $\Omega_\mathrm{m}=1$ is
the simplest model consistent with the spatial flatness expectation of
inflationary cosmology. However, Type Ia supernovae (SNe~Ia) at
redshift $z \simeq 0.5$ appear $\sim25\%$ fainter than expected in an
E-deS universe \citep{Riess:1998cb,Perlmutter:1998np}. Together with
measurements of galaxy clustering in the Two-degree Field survey
\citep{Efstathiou:2001cw} and of cosmic microwave background (CMB)
anisotropies by the Wilkinson Microwave Anisotropy Probe (WMAP)
\citep{Spergel:2003cb}, this has established an accelerating universe
with a dominant cosmological constant term (or other form of `dark
energy') which presumably reflects the present microphysical vacuum
state. This `concordance' $\Lambda$CDM cosmology (with $\Omega_\Lambda
\simeq 0.7$, $\Omega_\mathrm{m} \simeq 0.3$, $h \simeq 0.7$) has
passed a number of cosmological tests, including baryonic acoustic
oscillations \citep{Eisenstein:2005su} and measurements of mass
fluctuations from clusters and weak lensing
\citep[e.g.][]{Contaldi:2003hi}. Further observations of both SNe~Ia
\citep{Riess:2004nr,Astier:2005qq,WoodVasey:2007jb} and the WMAP
3-year results \citep{Spergel:2006hy} have continued to firm up the
model. However there is no physical basis for this model, in
particular there are two fundamental problems with the notion that the
universe is dominated by vacuum energy. The first is the notorious
fine-tuning problem of vacuum fluctuations in quantum field theory ---
the energy scale of the cosmological energy density is $\sim10^{-12}$
GeV, many orders of magnitude below the energy scale of $\sim10^2$ GeV
of the Standard Model of particle physics, not to mention the Planck
scale of $\sim10^{19}$ GeV \citep[see][]{Weinberg:1988cp}. The second
is the equally acute coincidence problem: since
$\rho_\Lambda/\rho_\mathrm{m}$ evolves as the cube of the cosmic scale
factor $a$, there is no reason to expect it to be of ${\cal O}(1)$
{\em today}, yet this is apparently the case. In fact what is actually
inferred from observations is {\em not} an energy density, just a
value of ${\cal O}(H_0^2)$ for the otherwise unconstrained $\Lambda$
term in the Friedmann equation. It has been suggested that this may
simply be an artifact of interpreting cosmological data in the
(oversimplified) framework of a perfectly homogeneous universe in
which $H_0 \sim 10^{-42} \mathrm{GeV} \sim (10^{28} \mathrm{cm})^{-1}$
is the {\em only} scale in the problem \citep{Sarkar:2007cx}.

In fact the WMAP results alone do {\em not} require dark energy if the
assumption of a scale-invariant primordial power spectrum is relaxed.
This assumption is worth examining given our present ignorance of the
physics behind inflation. We have demonstrated \citep{Hunt:2007dn}
that the temperature angular power spectrum of an E-deS universe with
$h \simeq 0.44$ matches the WMAP data well if the primordial power is
enhanced by $\sim 30\%$ in the region of the second and third acoustic peaks
(corresponding to spatial scales of $k \sim
0.01-0.1~h~\mathrm{Mpc}^{-1}$). This alternative model with {\em no}
dark energy actually has a slightly better $\chi^2$ for the fit to
WMAP-3 data than the `concordance power-law $\Lambda$CDM model' and,
inspite of having more parameters, has an {\em equal} value of the
Akaike information criterion used in model selection. Other E-deS
models with a broken power-law spectrum \citep{Blanchard:2003du} have
also been shown to fit the WMAP data.  Moreover, an E-deS universe can
fit measurements of the galaxy power spectrum if it includes a $\sim
10\%$ component of hot dark matter in the form of massive neutrinos of
mass $\sim 0.5$~eV \citep{Hunt:2007dn,Blanchard:2003du}.  Clearly the
main evidence for dark energy comes from the SNe~Ia Hubble diagram.

A mechanism that sets $\Lambda=0$ is arguably more plausible than one
which leads to the tiny energy density $\rho_\Lambda\simeq 10^{-47}$
GeV$^4$ associated with the concordance
cosmology.\footnote{`Quintessence' models, which attempt to address
  the coincidence problem, also {\em assume} that every other
  contribution to the vacuum energy cancels apart from that of the
  quintessence field.} If $\Lambda$ is indeed zero then perhaps some
effect fools us into wrongly deducing the existence of dark energy by
\emph{mimicking} a nonzero cosmological constant. It is natural to
connect this effect with inhomogenities since cosmic acceleration and
large scale nonlinear structure formation appear to have commenced
simultaneously. This approach offers the possibility of solving the
cosmological constant problems within the framework of general
relativity and keeps the introduction of new physics to a
minimum.\footnote{In models that seek to explain the observations
  through modifications of gravity, the relevant scale of $H_0^{-1}$
  has to be introduced {\em by hand}, just as in quintessence models
  the quintessence field has to be given a mass of order $H_0$ ---
  these are technically {\em unnatural} choices since this is an
  infrared scale for any microphysical theory.}  Several different
ways in which inhomogenities could potentially mimic dark energy have
been considered in the literature --- for reviews see
\citet{Celerier:2007jc,Buchert:2007ik,Enqvist:2007vb}. In an
inhomogeneous universe averaged quantities satisfy modified Friedmann
equations which contain extra terms corresponding to `backreaction'
since the operations of spatial averaging and time evolution do not
commute \citep{Buchert:1999er}. The backreaction terms depend upon the
variance of the local expansion rate and hence increase as
inhomogenities develop. Whether backreaction can indeed account for
the apparent cosmological acceleration is hotly debated and remains an
open question at present
\citep{Wetterich:2001kr,Ishibashi:2005sj,Vanderveld:2007cq,Wiltshire:2007fg,Khosravi:2007bq,Leith:2007ay,Behrend:2007mf,Rasanen:2008it,Paranjape:2008jc}.

Another possibility is that inhomogeneities affect light propagation
on large scales and cause the luminosity distance-redshift relation to
resemble that expected for an accelerating universe. This has been
investigated for a `Swiss-cheese' universe in which voids modelled by
patches of Lema\^{i}tre-Tolman-Bondi (LTB) space-time are distributed
throughout a homogenous background. However, the results seem to be
model dependent: some authors find the change in light propagation to
be negligible because of cancellation effects
\citep{Biswas:2007gi,Brouzakis:2007zi,Brouzakis:2008uw}, whereas
\citet{Marra:2007pm} claim it can partly mimic dark energy if the
voids have radius 250 Mpc \citep{Marra:2007gc}.
\citet{Mattsson:2007tj} has noted that observers may preferentially
choose sky regions with underdense foregrounds when studing distant
objects such as SNe~Ia, so the expansion rate along the line-of-sight
is then greater than average; such a selection effect he argues can
allow an inhomogeneous universe to fit the observations without dark
energy.
   
In this paper we are mainly interested in a `local void' (sometimes
referred to as ``Hubble bubble'') as an explanation for dark energy;
to prevent an excessive CMB dipole moment due to our peculiar velocity
we must be located near the centre of the void. An underdense void
expands faster than its surroundings, thus younger supernovae inside
the void would be observed to be receding more rapidly than older
supernovae outside the void. Under the assumption of homogeneity this
would lead to the mistaken conclusion that the expansion rate of the
Universe is accelerating, although both the void and the global
universe are actually decelerating. Henceforth we use the `Hubble
contrast' $\delta_H\equiv
\left(H_\mathrm{in}-H_\mathrm{out}\right)/H_\mathrm{out}$ to
characterise the void expansion rate, where $H_\mathrm{in}$ and
$H_\mathrm{out}$ are the Hubble parameters inside and outside the void
respectively. (Other authors have used the `jump' ${\mathcal
  J}\equiv\,H_\mathrm{in}/H_\mathrm{out} = 1 + \delta_H$ to
characterise the void.)  The reduced Hubble parameter $h$ is defined
as usual by $H_\mathrm{out} = 100h$ km $\mathrm{s}^{-1}$
$\mathrm{Mpc}^{-1}$ throughout.

The local void scenario has been investigated by several authors using
a variety of methods
\citep{Celerier:1999hp,Tomita:1999qn,Tomita:2000rf,Tomita:2000jj,Tomita:2001gh,Iguchi:2001sq,Tomita:2002df,Moffat:2005yx,Moffat:2005zx,Moffat:2005ii,Mansouri:2005rf,Vanderveld:2006rb,Garfinkle:2006sb,Chung:2006xh,Alnes:2005rw,Alnes:2006uk,Alexander:2007xx,Biswas:2006ub,Caldwell:2007yu,Clarkson:2007pz,Uzan:2008qp,GarciaBellido:2008nz,Clifton:2008hv,GarciaBellido:2008gd}.
In a series of papers, Tomita modelled the void as a open
Friedmann-Robertson-Walker (FRW) region joined by a singular mass
shell to a FRW background and found that a void with radius 200 Mpc
and $\delta_H=0.25$ fits the supernova Hubble diagram without dark
energy \citep{Tomita:2001gh}. \citet{Alnes:2005rw} showed that a LTB
region which reduces to a E-deS cosmology with $h=0.51$ at a radius of
1.4~Gpc with $\delta_H=0.27$ can match both the supernova data and
the location of the first acoustic peak in the
CMB. \citet{Alexander:2007xx} attempted to find the smallest possible
void consistent with the current supernova results --- their LTB-based
`minimal void' model has a radius of 350 Mpc and ${\mathcal J} \simeq
1.2$ i.e.  $\delta_H \simeq 0.2$; a void of similiar size but with
$\delta_H=0.3$ had been discussed earlier
\citep{Biswas:2006ub}. Unfortunately, since this model is equivalent
to an E-deS universe with $h=0.44$ {\em outside} the void where the
Sloan Digital Sky Survey (SDSS) luminous red galaxies lie, as it
stands it is unable to fit the measurements of the baryonic acoustic
oscillation (BAO) peak at $z\sim0.35$ \citep{Blanchard:2005ev}.  LTB
models of much larger voids were considered by
\citet{GarciaBellido:2008nz} (with radii of 2.3 Gpc and 2.5 Gpc and
Hubble contrasts of 0.18 and 0.30 respectively) and it was
demonstrated they can fit the supernova data, BAO data and the
location of the first CMB peak. \citet{Clifton:2008hv} found the best
fit to the SNe Ia data for a void of radius $1.3 \pm 0.2$ Gpc and an
underdensity of about $70\%$ at the centre and \citet{Bolejko:2008cm}
confirmed that such a void provides an excellent fit to the latest
`Union dataset' \citep{Kowalski:2008ez}. Moreover \citet{Inoue:2006rd}
have shown that the unexpected alignment of the low multipoles in the
CMB anisotropy can be attributed to the existence of a local void of
radius $300$~$h^{-1}\,\mathrm{Mpc}$. These authors also suggested that
the anomalous `cold spot' in the WMAP southern sky is due to a similar
void at $z \sim 1$ and some evidence for this has emerged subsequently
\citep{Rudnick:2007kw}.  Recently, a large number of voids of varying
sizes have been identified in the SDSS Luminous Red Galaxy (LRG)
catalog in a search for the late integrated Sachs-Wolfe (ISW) effect
due to dark energy \citep{Granett:2008ju}.

How likely is the existence of such huge voids according to standard
theories of structure formation? Statistical measures of the void
distribution such as the void probability function and underdense
probability function have been estimated from the 2dfGRS, SDSS and
DEEP2 galaxy redshift surveys
\citep{Hoyle:2003hc,Croton:2004ac,Patiri:2005ys,Conroy:2005cx,Tikhonov:2006it,Tinker:2007zi,Tikhonov:2007di,vonBendaBeckmann:2007wt}.
Void probability statistics have also been examined theoretically
using analytical methods
\citep{Sheth:2003py,Furlanetto:2005cc,Shandarin:2005ea} and N-body
simulations
\citep{Little:1993fe,Schmidt:2000md,ArbabiBidgoli:2001jp,Benson:2002tq,Padilla:2005ea}.
However such studies have been restricted to voids with radii of 10-30
Mpc. The scales of the large voids we are considering lie in the
linear regime where the variance of the Hubble contrast is directly
related to the matter power spectrum
$\mathcal{P}_\mathrm{m}\left(k\right)$. It has been noted (using
results from \citet{Turner:1992yi}) that above 100 Mpc linear theory
predictions agree well with N-body simulation results, although on
smaller scales the Hubble contrast is underestimated due to non-linear
effects \citep{Shi:1995nq}. Applying linear theory and using the
measured CMB dipole velocity, \citet{Wang:1997tp} obtained the
model-independent result $\langle\delta_H\rangle^{1/2}_R <
10.5\,h^{-1}\,\mathrm{Mpc}/R$ in a sphere of radius $R$. (This ought
to be an acceptable procedure up to scales of order 800 $h^{-1}$~Mpc
--- on larger scales, relativistic corrections become increasingly
important.) In this paper we update these results by determining the
probability distribution of $\delta_H$ and the density contrast on
various scales using constraints on
$\mathcal{P}_\mathrm{m}\left(k\right)$ from WMAP 5-year data
\citep{Komatsu:2008hk} and the SDSS galaxy power spectrum
\citep{Tegmark:2003uf}.  We find that even the `minimal local void' is
{\em extremely} unlikely if the primordial density perturbation is
indeed gaussian as is usually assumed and the other LTB model voids
even less so. However by the same token, the ISW effect due to the
voids seen in the SDSS LRG survey \citep{Granett:2008ju} appears to be
too strong. Moreover, observed large-scale peculiar velocities appear
to be much higher than expected
\citep{Kashlinsky:2008ut,Watkins:2008hf}. It would appear that the
standard model of structure formation itself needs reexamination hence
the existence of a large local void cannot be dismissed on these
grounds.

\section{Models}

We study variations of the Hubble parameter in the context of two
different cosmological models, both of which fit the WMAP and SDSS
data but have different amounts of power on spatial scales of ${\cal
  O}(100)$ Mpc. The intention is to examine whether previous
conclusions concerning the magnitude of such variations
\citep{Wang:1997tp} can be circumvented in an unorthodox model.

Our first model is the standard $\Lambda$CDM concordance model with a
power-law primordial power spectrum. The spectral index and amplitude
$\mathcal{P_R}$ of the comoving curvature perturbation spectrum are
evaluated at a pivot point of $k = 0.05\,\mathrm{Mpc}^{-1}$. The
second model is dubbed the `CHDM bump model' since it has both cold
and hot dark matter and a `bump' in the primordial spectrum. It was
developed by us \citep{Hunt:2007dn} based upon the supergravity
multiple inflation scenario in which `flat direction' fields undergo
gauge symmetry-breaking phase transitions during inflation triggered
by the fall in temperature \citep{Adams:1997de,Hunt:2004vt}. Each flat
direction $\psi$ has a gravitational strength coupling to the
inflaton $\phi$, giving a contribution to the potential of the form
$V\subset\frac{1}{2}\lambda\phi^2\psi^2$. The flat directions are
lifted by supergravity corrections and non-renormalisable
superpotential terms.  Thus when a phase transition occurs the flat
direction evolves rapidly from the origin where it was trapped by
thermal effects to the global minimum of the potential. Each phase
transition changes the effective inflaton mass from $m_\phi^2$ to
$m_\phi^2-\lambda\left<\psi\right>^2$. Since the primordial power
spectrum is very sensitive to the inflaton mass this can introduce
features into the spectrum. We showed that two
flat directions $\psi_1$ and $\psi_2$ which cause successive phase
transitions about 2 e-folds apart and create a small bump in the
power spectrum centred on $k\simeq0.03\, h\,
\mathrm{Mpc}^{-1}$, allow an E-deS model with $h=0.44$ to fit the WMAP
data \citep{Hunt:2007dn}. The effective scalar potential is:
\begin{equation}
V \left(\phi, \psi_1, \psi_2\right) =\left\{\begin{array}{ll}
V_0 - \frac{1}{2}m^2 \phi^2, & t < t_1,\\
V_0 - \frac{1}{2}m^2\phi^2 - \frac{1}{2}\mu_1^2\psi_1^2\\
  {} + \frac{1}{2}\lambda_1\phi^2\psi_1^2
 + \frac{\gamma_1}{M_\mathrm{P}^{n_1 - 4}}\psi^{n_1}, 
& t_2 \geq t \geq t_1,\\
V_0 - \frac{1}{2}m^2\phi^2 - \frac{1}{2}\mu_1^2\psi_1^2\\ 
{} + \frac{1}{2}\lambda_1\phi^2\psi_1^2
+\frac{\gamma_1}{M_{P}^{n_1 -4}}\psi^{n_1}\\
{} - \frac{1}{2}\mu_2^2\psi_2^2 - \frac{1}{2}\lambda_2\phi^2\psi_2^2\\
{} + \frac{\gamma_2}{M_\mathrm{P}^{n_2 - 4}}\psi^{n_2}, & t \geq t_2.
\end{array}\right.
\end{equation}
Here $t_1$ and $t_2$ are the times at which the first and second phase
transitions begin, $\lambda_1$ and $\lambda_2$ are the couplings
between $\phi$ and the flat directions, $\gamma_1$ and $\gamma_2$ are
the co-efficients of the non-renormalisable terms of order $n_1$ and
$n_2$, and $V_0$ is a constant which dominates the potential. In the
slow-roll approximation the height of the bump is
$\mathcal{P_R}^{(1)}$, and the amplitude of the primordial
perturbation spectrum to the left and right of the bump is
$\mathcal{P_R}^{(0)}$ and $\mathcal{P_R}^{(2)}$ respectively, where
\begin{eqnarray}
\mathcal{P_R}^{(0)} & = &
\frac{9H^6}{4\pi^2 m^4\phi_0^2},\\ 
\mathcal{P_R}^{(1)} & = &
\frac{\mathcal{P}_{\mathcal{R}}^{\left(0\right)}}{\left(1-\Delta
  m_{1}^{2}\right)^{2}}, \\ 
\mathcal{P_R}^{(2)} & = &
 \frac{\mathcal{P_R}^{(0)}}{\left(1 - \Delta m_{1}^{2} + 
 \Delta m_{2}^2\right)^2}\ .
\label{ampl}
\end{eqnarray}
Here $\phi_0$ is the initial value of $\phi$ and
\begin{eqnarray}
\Delta m_1^2 & = & \frac{\lambda_1}{m^2}\left(\frac{\mu_1^2 M_\mathrm{P}^{n_1-4}}
{n_1\gamma_1}\right)^{2/\left(n_1-2\right)}, \\
\Delta m_2^2 & = & \frac{\lambda_2}{m^2}\left(\frac{\mu_2^2 M_\mathrm{P}^{n_2-4}}
{n_2\gamma_2}\right)^{2/\left(n_2-2\right)},
\end{eqnarray}  
are the fractional changes in the inflaton mass-squared due to the
phase transitions. The bump lies approximately between the wavenumbers
$k_1$ and $k_2$ where $k_2=k_1 e^{H\left(t_2-t_1\right)}$. In this
paper we set $\gamma_1$ and $\gamma_2$ equal to unity, $m^2=0.005H^2$,
$\phi_0=0.01M_\mathrm{P}$, $\mu_1^2 =\mu_2^2 = 3H^2$ and
$\lambda_1=\lambda_2=H^2/M_\mathrm{P}^2$ throughout as in our earlier
work \citep{Hunt:2007dn}. In fitting to the WMAP-5 data we also
consider continuous (non-integral) values of $n_1$ and $n_2$ to
determine whether a different shape of the `bump' gives a better fit,
keeping in mind that its physical origin may be different from
multiple inflation
\citep{Chung:1999ve,Lesgourgues:1999uc,Kaloper:2003nv,
  Easther:2001fi,Wang:2002hf,Gong:2005jr,Ashoorioon:2006wc,Bean:2008na}.

A pure cold dark matter (CDM) model exhibits excessive galaxy
clustering on small scales. Therefore it is necessary to include a hot
dark matter (HDM) component which suppresses structure formation below
the free-streaming scale. We obtain a good match to the shape of the
SDSS galaxy power spectrum with 3 neutrino species of mass $\sim 0.5$
eV. Hence the CHDM bump model has
$\Omega_\mathrm{b}\simeq0.1,\,\Omega_\nu\simeq0.1,\,\Omega_\mathrm{c}\simeq0.8$
\citep{Hunt:2007dn}.

\section{The Data Sets}

We fit to the WMAP 5-year \citep{Nolta:2008ih} temperature-temperature
(TT), temperature-electric polarisation (TE), and electric-electric
polarisation (EE) spectra. Compared to the WMAP-3 results, the WMAP-5
measurement of the TT spectrum is $\sim2.5\%$ higher in the region of
the acoustic peaks due to the revised beam transfer functions, and the
third acoustic peak is determined more accurately. Polarisation
measurements are improved by the use of data from an additional
waveband.

We also fit the linear matter power spectrum $\mathcal{P}_\mathrm{m}
(k)$ to the measurement of the real space galaxy power spectrum
$\mathcal{P}_\mathrm{g}(k)$ in the SDSS
\citep{Tegmark:2003uf}.

\section{Method}

The Hubble contrast $\delta_H$ smoothed over a sphere of radius $R$ is
\citep{Shi:1995nq}
\begin{equation}
\delta_H({\bf x})^R = \int d^3{\bf y} \,\frac{{\bf v}({\bf y})}
{H_\mathrm{out}} \cdot \frac{ {\bf y}-{\bf x} }{|{\bf y}-{\bf x}|^2} \, 
W_R({\bf y}-{\bf x}),
\end{equation}  
where $\bf v$ is the peculiar velocity field and $W_R$ is the `top
hat' window function,
\begin{equation}
W_R({\bf x})= \left\{ \begin{array}{ll} 3/(4\pi R^3), &
\qquad |{\bf x}| \leq R, \\
0, & \qquad |{\bf x}| > R. \end{array} \right.
\end{equation}
Using linear perturbation theory \citep{Peebles:1994xt} it can be shown
that the variance of $\delta_H$ is related to the matter power
spectrum as \citep{Wang:1997tp}
\begin{equation}
\left\langle \delta_H^2\right\rangle _R=\frac{f^2}{2\pi^2}\int_0^\infty
\mathrm{d}k\,k^2 \mathcal{P}_\mathrm{m}\left(k\right)W^2_H\left(kR\right).
\label{var1}
\end{equation}
Here the window function $W_H$ is 
\begin{equation}
W_H\left(kR\right)=\frac{3}{k^3R^3}\left(\sin kR-\int_0^{kR} 
\mathrm{d}y\,\frac{\sin y}{y}\right),
\label{wh}
\end{equation}
and the dimensionless linear growth rate $f$ for a $\Lambda$CDM
universe can be approximated by
\citep{Lahav:1991wc,Hamilton:2000tk}\footnote{\citet{Hamilton:2000tk}
  emphasized that the power-law exponent is 4/7 for a high density
  universe, so we have corrected the previous formula from
  \citet{Lahav:1991wc} accordingly.}
\begin{equation}
f\left(\Omega_\mathrm{m},\Omega_\Lambda\right)\simeq
\Omega_\mathrm{m}^{4/7}+\frac{\Omega_\Lambda}{70}
\left(1+\frac{\Omega_\mathrm{m}}{2}\right).
\end{equation}

Similarly, the variance of the density contrast
$\delta\equiv\left(\rho_\mathrm{in}-\rho_\mathrm{out}\right)/\rho_\mathrm{out}$
in a sphere of radius $R$ is
\begin{equation}
\left\langle \delta^2\right\rangle _R=\frac{1}{2\pi^2}\int_0^\infty
\mathrm{d}k\,k^2 \mathcal{P}_\mathrm{m}\left(k\right)W^2\left(kR\right),
\label{var2}
\end{equation}
where the window function 
\begin{equation}
W\left(kR\right)=\frac{3}{k^3 R^3}\left(\sin kR-kR\cos kR\right),
\end{equation}
is the Fourier transform of $W_R$. 

The variance of the peculiar velocity is given by
\begin{equation}
  \left\langle {v}^2\right\rangle _R = 
  \frac{f^2 H_\mathrm{out}^2}{2 \pi^2}\int_0^\infty
  \mathrm{d}k\,\mathcal{P}_\mathrm{m}\left(k\right)W^2\left(kR\right).
\label{var3}
\end{equation} 

Finally we also consider $\Omega_\mathrm{in}=8\pi
G\rho_\mathrm{in}/H_\mathrm{in}^2$, the ratio of the matter density to
the critical density as measured locally by an observer inside the
void \citep{Wang:1997tp}. The variance of the perturbation
$\delta_\Omega \equiv
\left(\Omega_\mathrm{in}-\Omega_\mathrm{m}\right)/\Omega_\mathrm{m}$
is then
\begin{equation}
\left\langle \delta_\Omega^2\right\rangle _R=\frac{1}{2\pi^2}\int_0^\infty
\mathrm{d}k\,k^2 \mathcal{P}_\mathrm{m}\left(k\right)W^2_\Omega\left(kR\right),
\label{var4}
\end{equation}
where
\begin{eqnarray}
W_\Omega\left(kR\right)& = & \frac{3}{k^3 R^3}\bigg[\left(2f-1\right)
\sin kR\nonumber\\& & \left.{}+kR\cos kR+2f\int_0^{kR} 
\mathrm{d}y\,\frac{\sin y}{y}\right].
\end{eqnarray}

We use the Monte Carlo Markov Chain (MCMC) approach to cosmological
parameter estimation, which is a method for drawing samples from the
posterior distribution
$P\left(\mbox{\boldmath$\varpi$}|\mathrm{data}\right)$ of the
cosmological parameters $\mbox{\boldmath$\varpi$}$, given the data.
For a discussion of the MCMC likelihood analysis see Appendix B of
\citet{Hunt:2007dn}. Given $n$ samples
$\mbox{\boldmath$\varpi$}^{\left(i\right)}$ the best estimate for the
distribution is
\begin{equation}
P\left(\mbox{\boldmath$\varpi$}\right|\textrm{data})
\simeq\frac{1}{n}\sum_{i=1}^{n}\delta^D\left(\mbox{\boldmath$\varpi$}
-\mbox{\boldmath$\varpi$}^{\left(i\right)}\right),
\label{mcmc1}
\end{equation}
where $\delta^D$ is the Dirac delta function.  The CMB angular power
spectrum and the matter power spectrum (corrected for non-linear
evolution using the `Halofit' \citep{Smith:2002dz} procedure) of each
model are calculated using a modified version of the {\sc
  camb}\footnote{http://camb.info} cosmological Boltzmann code
\citep{Lewis:1999bs} following the approach of \citet{Hunt:2007dn}.
While the temperature of the CMB monopole would be affected if we are
located near the centre of a spherically symmetric void, secondary
anisotropies due to the void are expected to decay rapidly for higher
multipoles \citep{Alexander:2007xx}. We therefore neglect the possible
effects of the void on the angular power spectra since these are
important only at low multipoles where the cosmic variance is large.
The integrals for the variances in eqs.(\ref{var1},
\ref{var2}-\ref{var4}) were evaluated numerically.  Care was taken to
ensure the precise values of the integration limits did not affect the
results.  We compute $f$ numerically using the {\sc grow}$\lambda$
software package \citep{Hamilton:2000tk}.  We use a version of the
{\sc cosmomc}\footnote{http://cosmologist.info/cosmomc/} package
\citep{Lewis:2002ah} modified to include $\left\langle
  \delta_H^2\right\rangle _R^{1/2}$, $\left\langle
  \delta^2\right\rangle _R^{1/2}$, $\left\langle {v}^2\right\rangle
_R^{1/2}$ and $\left\langle \delta_\Omega^2\right\rangle _R^{1/2}$ for
8 values of $R$ as additional derived parameters determined from the
base Monte Carlo parameters. It follows from eq.(\ref{mcmc1}) that 1-D
marginalised distributions of these quantities for each $R$ value are
obtained by plotting histograms of the samples. The probability
distribution $P\left(\delta_{H}\right|\textrm{data})_R$ of $\delta_H$
on the scale $R$ given the data can be written as
\begin{equation}
P\left(\delta_{H}\right|\textrm{data})_R=
\int P\left(\delta_{H}\right|\mbox{\boldmath$\varpi$})_{R}P
\left(\mbox{\boldmath$\varpi$}\right|\textrm{data})
\,\mathrm{d}\mbox{\boldmath$\varpi$}.
\end{equation}
Using eq.(\ref{mcmc1}) this is approximated by
\begin{equation}
  P\left(\delta_{H}\right|\textrm{data})_R=
  \frac{1}{n}\sum_{i=1}^{n}P\left(\delta_{H}\right| 
  \mbox{\boldmath$\varpi$}^{(i)})_{R},
\label{sum}
\end{equation}
where
\begin{equation}
P\left(\delta_{H}\right|\mbox{\boldmath$\varpi$})_{R} = 
\frac{1}{\sqrt{2\pi\left\langle \delta_{H}^{2}\right
\rangle _{R}}}\exp\left(-\frac{\delta_{H}^{2}}
{2\left\langle \delta_{H}^{2}\right\rangle _{R}}\right).
\end{equation}
We calculate the probability distribution $P\left(\delta
\right|\textrm{data})_R$ in same way.

Flat priors are used on the parameters listed in Table
\ref{tab:priors}. Here $\theta$ is the ratio of the sound horizon to
the angular diameter distance to last scattering (multiplied by 100),
$\tau$ is the optical depth (due to reionisation) to the last
scattering surface, and $f_\nu \equiv\Omega_{\nu}/\Omega_{\rm d}$ is
the fraction of dark matter in the form of neutrinos, where the total
dark matter density is $\Omega_{\rm d} \equiv \Omega_{\rm c} +
\Omega_{\nu}$.  We assume the chains have converged when the
Gelman-Rubin `R' statistic falls below 1.02. We evaluate the sum in
eq.(\ref{sum}) when post-processing the chains.

\begin{table*}
\begin{minipage}{150mm}
\caption{\label{tab:priors} 
  The priors adopted on the base Monte Carlo parameters of the various models,
  as well as on the derived parameters: the
  Hubble constant and the age of the Universe.}
\begin{center}
\begin{tabular}{||l|l|l|l|l|l|l||}
  \hline
  \hline
  Parameter & \multicolumn{6}{c||}{Model} \\
  \cline{2-7}
  & \multicolumn{2}{l|}{$\Lambda$CDM power-law} & \multicolumn{2}{l|}{CHDM bump with} & \multicolumn{2}{l||}{CHDM bump with} \\
  & \multicolumn{2}{l|}{} & \multicolumn{2}{l|}{$n_1=12$, $n_2=13$} & \multicolumn{2}{l||}{$n_1$, $n_2$ continuous} \\
  \cline{2-7}
  & Lower limit & Upper limit & Lower limit & Upper limit & Lower limit & Upper limit \\
  \hline
  \hline
  $\Omega_\mathrm{b}h^2$ & $0.005$ & $0.1$ & $0.005$ & $0.1$ & $0.005$ & $0.1$ \\
  \hline
  $\Omega_\mathrm{c}h^2$ & $0.01$ & $0.99$ & & & & \\
  \hline
  $\theta$ & $0.5$ & $10.0$ & $0.5$ & $10.0$ & $0.5$ & $10.0$ \\
  \hline
  $\tau$ & $0.01$ & $0.8$ & $0.01$ & $0.8$ & $0.01$ & $0.8$ \\
  \hline
  $f_\nu$ & & & $0.01$ & $0.3$ & $0.01$ & $0.3$ \\
  \hline
  $n_\mathrm{s}$ & $0.5$ & $1.5$ & & & & \\
  \hline
  $10^{4}k_{1}/\mathrm{Mpc}^{-1}$ & & & 0.01 & 600 & 0.01 & 600 \\
  \hline
  $10^{4}k_{2}/\mathrm{Mpc}^{-1}$ & & & 0.01 & 800 & 0.01 & 1100 \\
  \hline
  $\ln\left(10^{10}\mathcal{P_R}\right)$ & $2.7$ & $4.0$ & & & & \\
  \hline
  $\ln\left(10^{10}\mathcal{P_R}^{(0)}\right)$ & & & $2.0$ & $6.0$ & $2.0$ & $6.0$ \\
  \hline
  $\ln\left(10^{10}\mathcal{P_R}^{(1)}\right)$ & & & & & $2.0$ & $6.0$ \\
  \hline
  $\ln\left(10^{10}\mathcal{P_R}^{(2)}\right)$ & & & & & $2.0$ & $6.0$ \\
  \hline
  $h$ & $0.4$ & $1.0$ & $0.1$ & $1.0$ & $0.1$ & $1.0$ \\
  \hline
  Age/Gyr & $10.0$ & $20.0$ &  $10.0$ & $20.0$ & $10.0$ & $20.0$ \\
  \hline
  \hline
\end{tabular}
\end{center}
\end{minipage}
\end{table*}

\section{Results}

The mean values of the marginalised cosmological parameters together
with their $68\%$ confidence limits are listed in Table
\ref{tab:parameters}. As in our previous work \citep{Hunt:2007dn} we
also list the value of the Akaike information criterion (AIC) relative
to the $\Lambda$CDM power-law model. Recall that the AIC is defined as
$\mathrm{AIC}\equiv -2\ln\mathcal{L}_{\mathrm{max}} + 2N$
\citep{Akaike:1974} where ${\cal L}_\mathrm{max}$ is the maximum
likelihood and $N$ the number of parameters. It is a commonly used
guide for judging whether additional parameters are warranted given
the increased model complexity, and quantifies the compromise between
improving the fit and adding extra parameters.

\begin{figure*}
\begin{minipage}{150mm}
\includegraphics*[angle=0,scale=0.5]{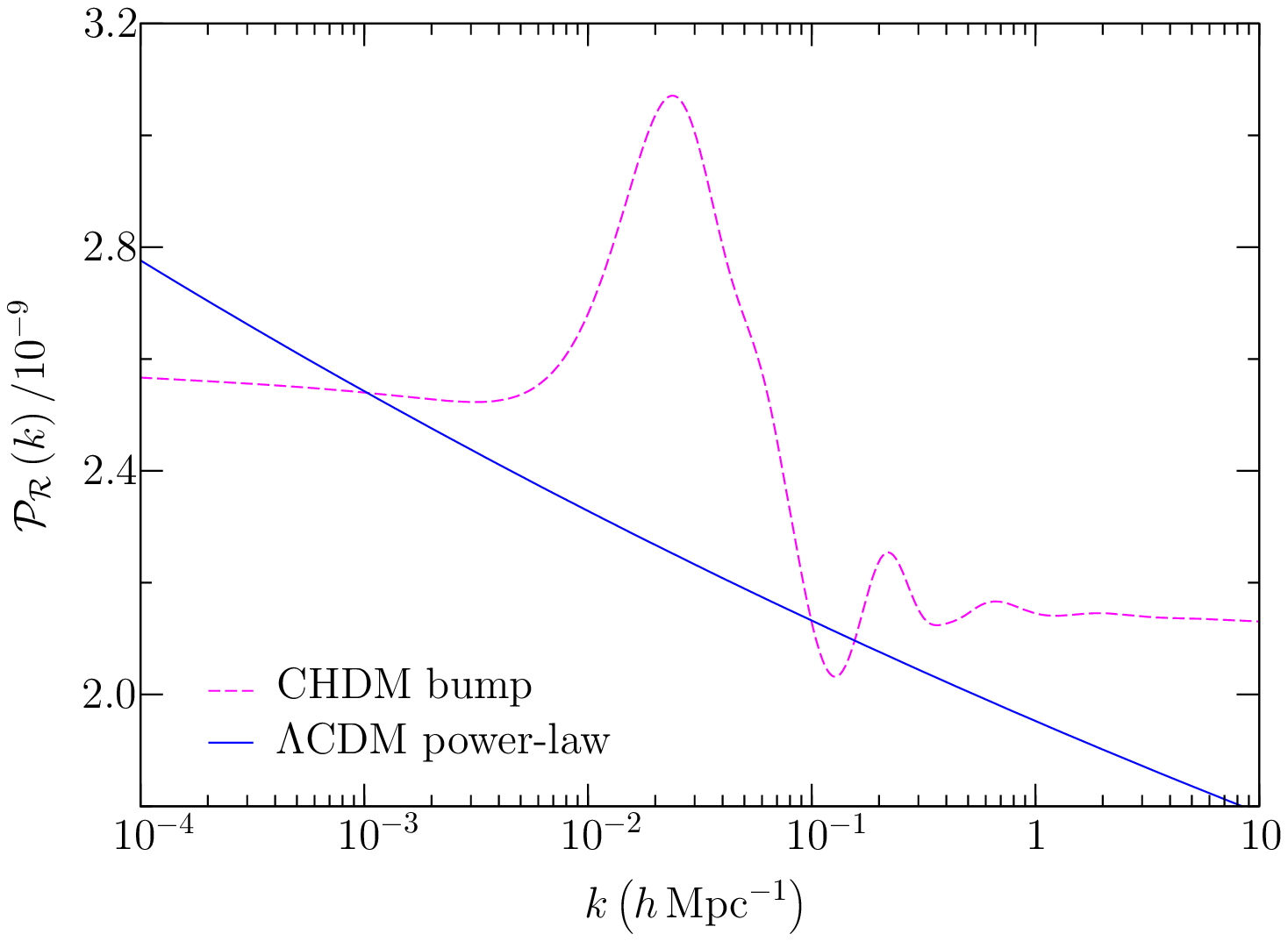}
\includegraphics*[angle=0,scale=0.5]{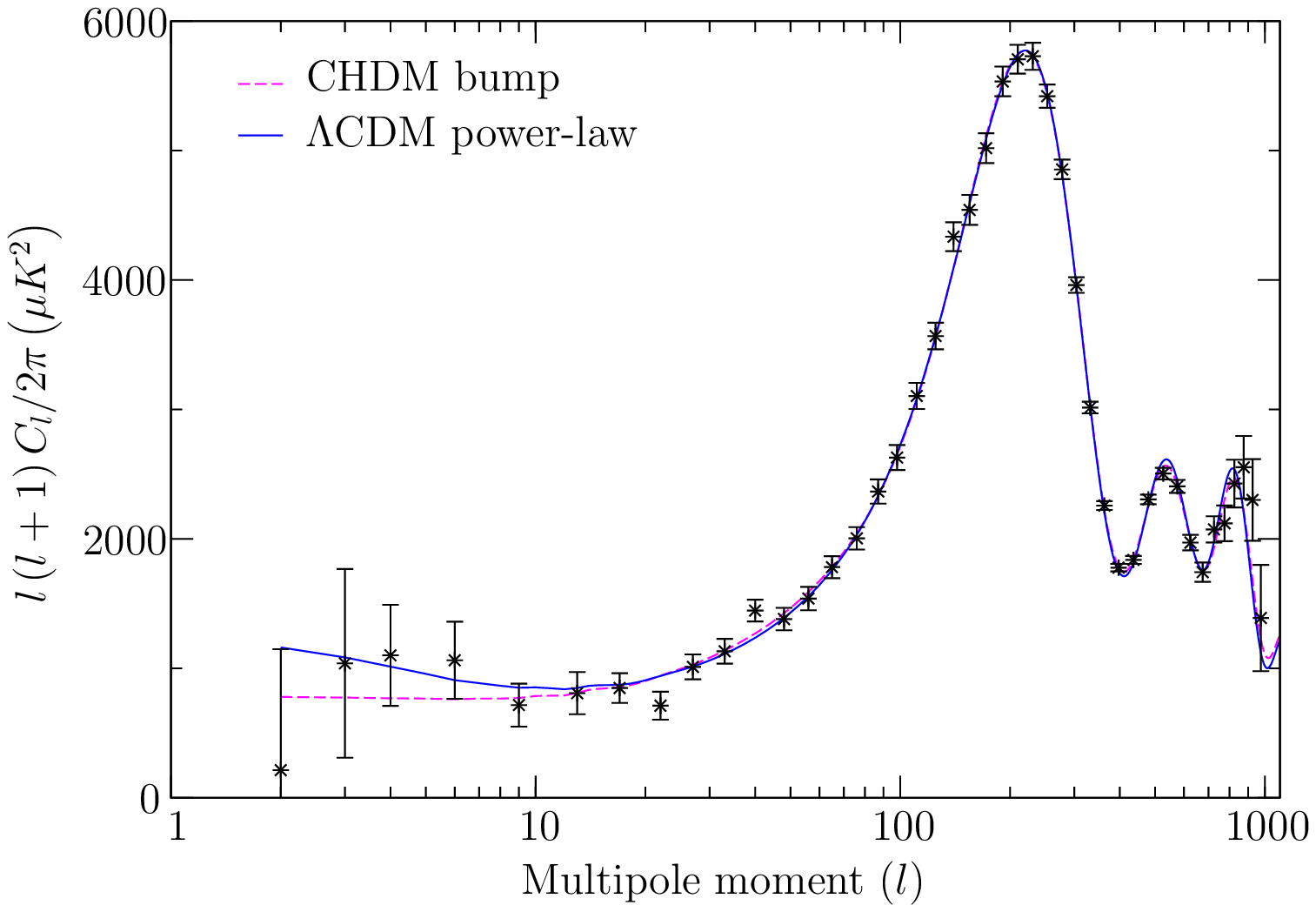}
\includegraphics*[angle=0,scale=0.5]{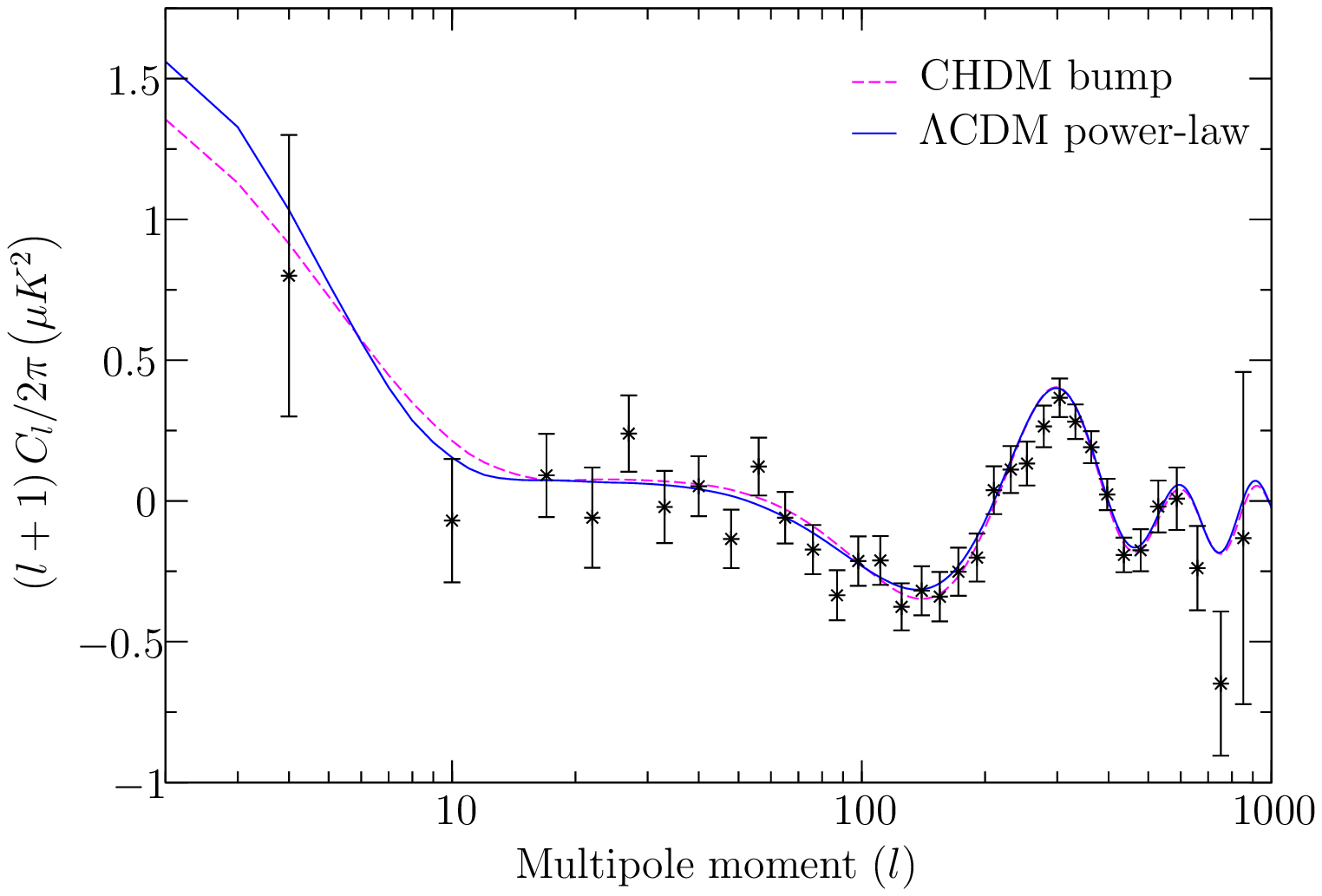}
\includegraphics*[angle=0,scale=0.5]{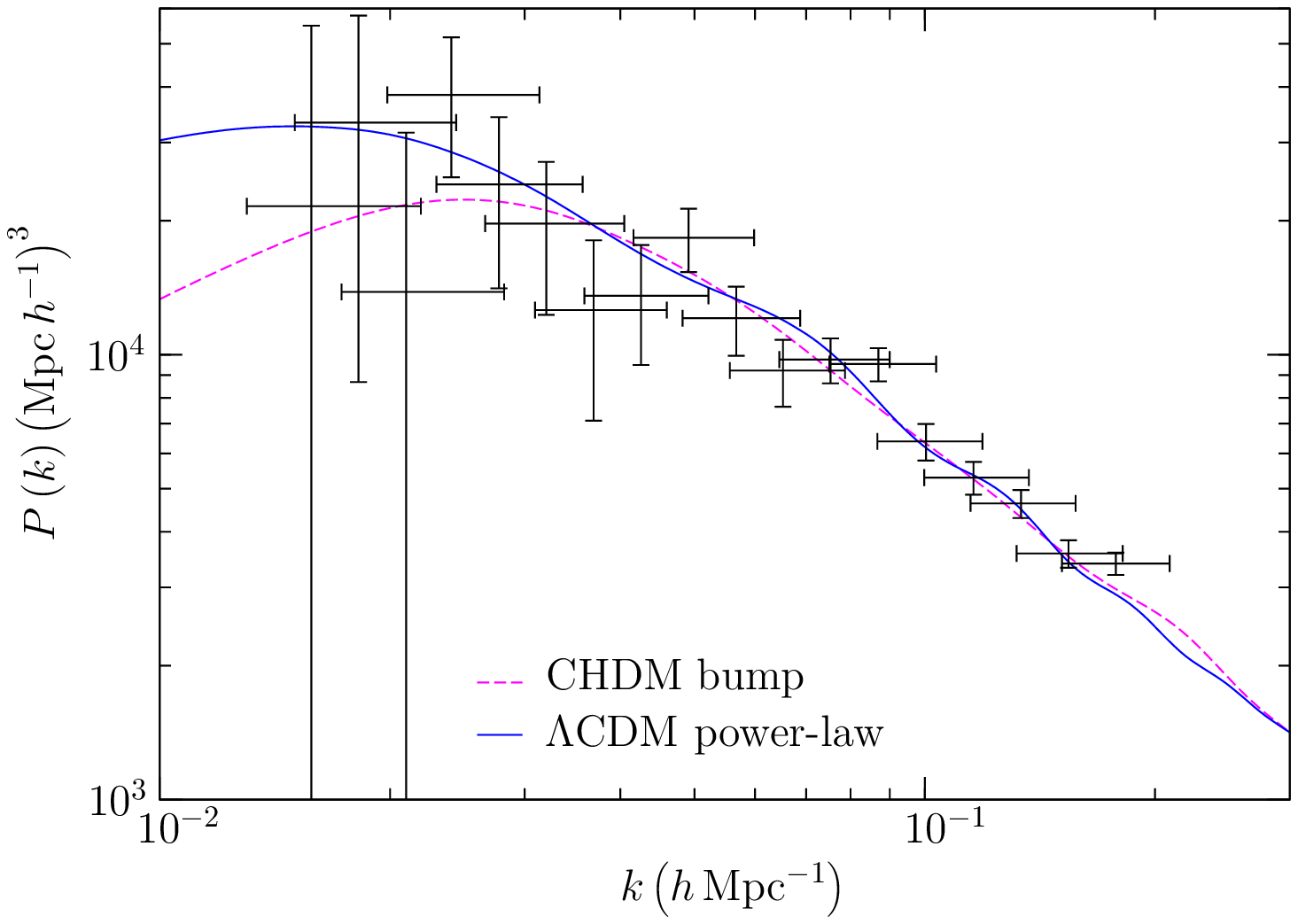}
\caption{\label{wmapfits} The top left panel shows the primordial
  perturbation spectrum for the CHDM bump model (with $n_1=12$ and
  $n_2=13$) and for the $\Lambda$CDM power-law model with $n_\mathrm{s}
  \simeq 0.96$. The top right and bottom left panels show the
  best-fits for both models to the WMAP-5 TT and TE spectra, while the
  bottom right panel shows the best-fits to the SDSS galaxy power
  spectrum.}
\end{minipage}
\end{figure*}

The CHDM `bump' model with $n_1=12$ and $n_2=13$ has a $\chi^2$ equal
to the $\Lambda$CDM power-law model. Allowing $n_1$ and $n_2$ to vary
freely further improves the fit to the data with the consequence that
the CHDM model with $n_1$ and $n_2$ continuous is {\em favoured} over
the $\Lambda$CDM model according to the AIC. The primordial power
spectrum of the models is shown in Fig.\ref{wmapfits} together with
the fit to the WMAP TT and TE spectra and the SDSS galaxy power
spectrum.

The uncertainties of the derived parameters are smaller compared to
those derived from the WMAP 3-year results, as would be expected for
higher quality data. For example, the optical depth due to
reionisation for the CHDM model with continuous $n_1$ and $n_2$ has
gone from $\tau=0.075_{-0.012}^{+0.012}$ to
$\tau=0.0771_{-0.0083}^{+0.0073}$ due to the more accurate
polarisation measurements. The shape of the `bump' in the primordial
power spectrum for the CHDM model with continuous $n_1$ and $n_2$ is
slightly changed by the new data.  Although the quantity
$\ln\left(10^{10}\mathcal{P_R}^{(0)}\right)$ is almost unaltered,
$\ln\left(10^{10}\mathcal{P_R}^{(1)}\right)$ has increased slightly
from a 3-year value of $3.429_{-0.049}^{+0.048}$ to a 5-year value of
$3.462_{-0.036}^{+0.036}$ because of the increased amplitude of the TT
spectrum for multipoles $\ell>200$. Due to the increased height of the
third acoustic peak, $\ln\left(10^{10}\mathcal{P_R}^{(2)}\right)$ has
increased from $3.091_{-0.067}^{+0.071}$ to $3.183_{-0.041}^{+0.043}$
and $10^{4}k_{2}$ fallen from $585_{-82}^{+36}\;\mathrm{Mpc}^{-1}$ to
$500_{-54}^{+21}\;\mathrm{Mpc}^{-1}$. The increased amplitude of the
primordial power spectrum on small scales has raised $\sigma_8$ from a
value of $0.662_{-0.064}^{+0.063}$ to $0.700_{-0.098}^{+0.098}$.

\begin{figure*}
\begin{minipage}{\textwidth}
\centering
\begin{tabular}{cc}
\includegraphics*[scale=0.55]{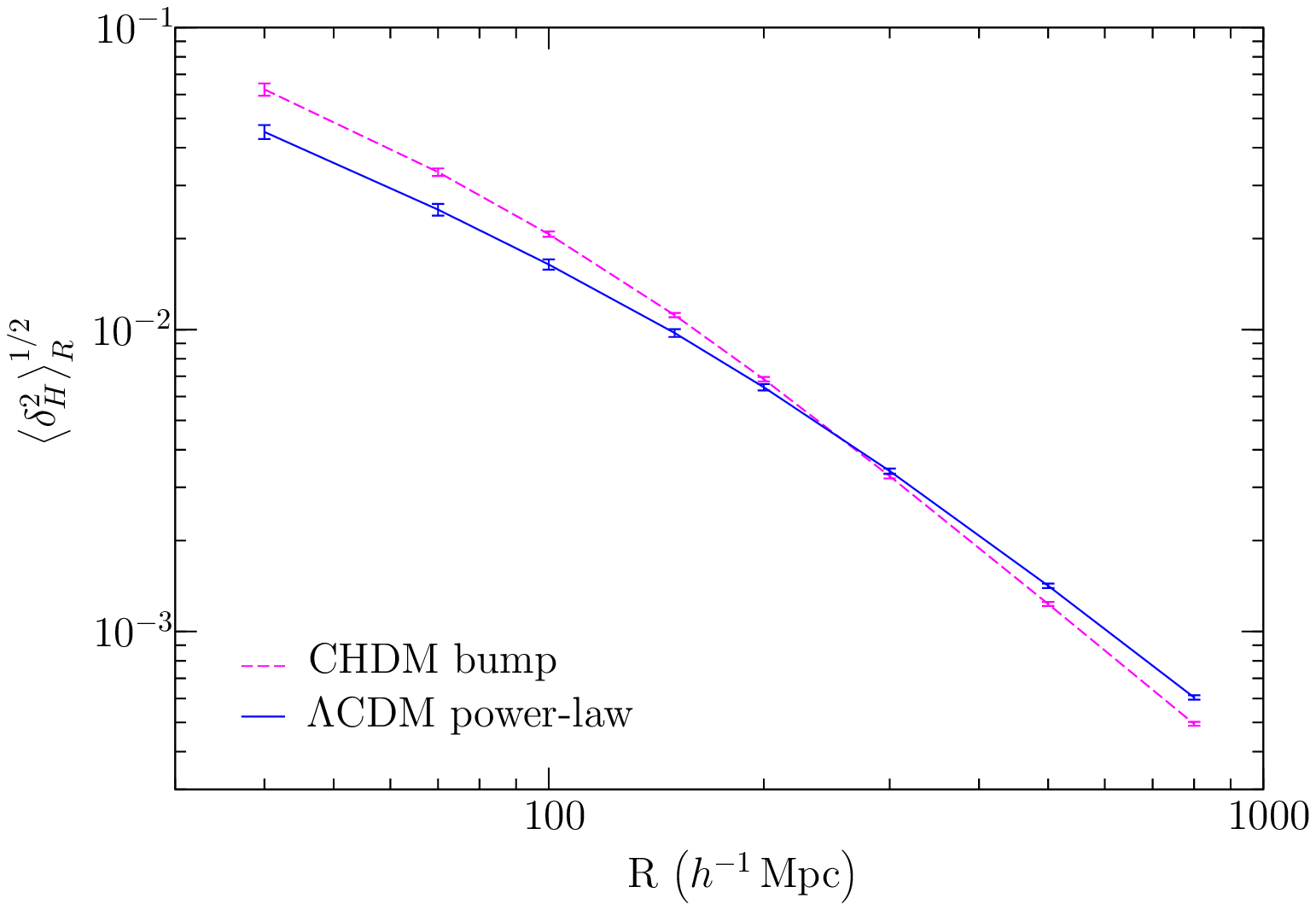} &
\includegraphics*[scale=0.55]{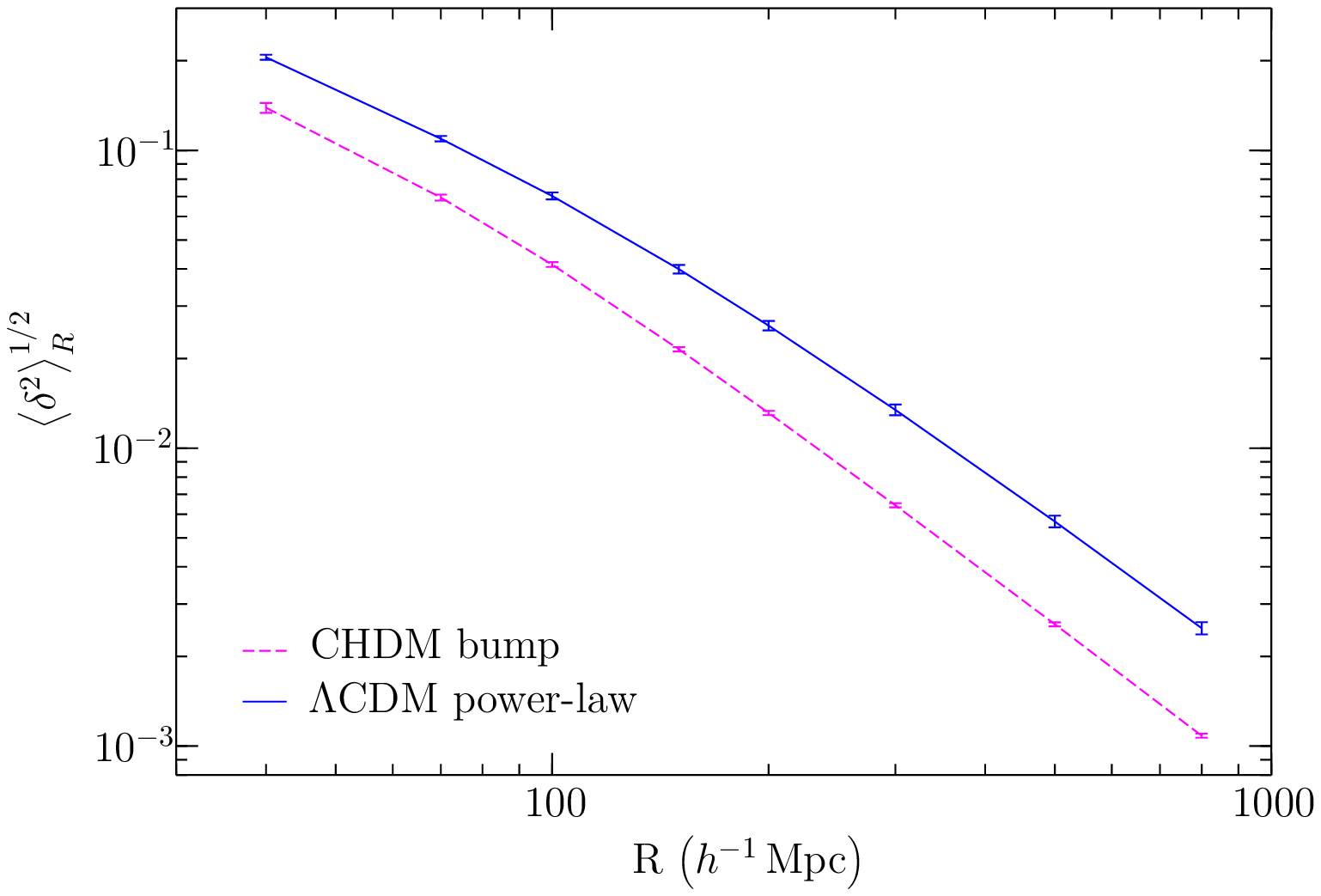} \\
\includegraphics*[scale=0.55]{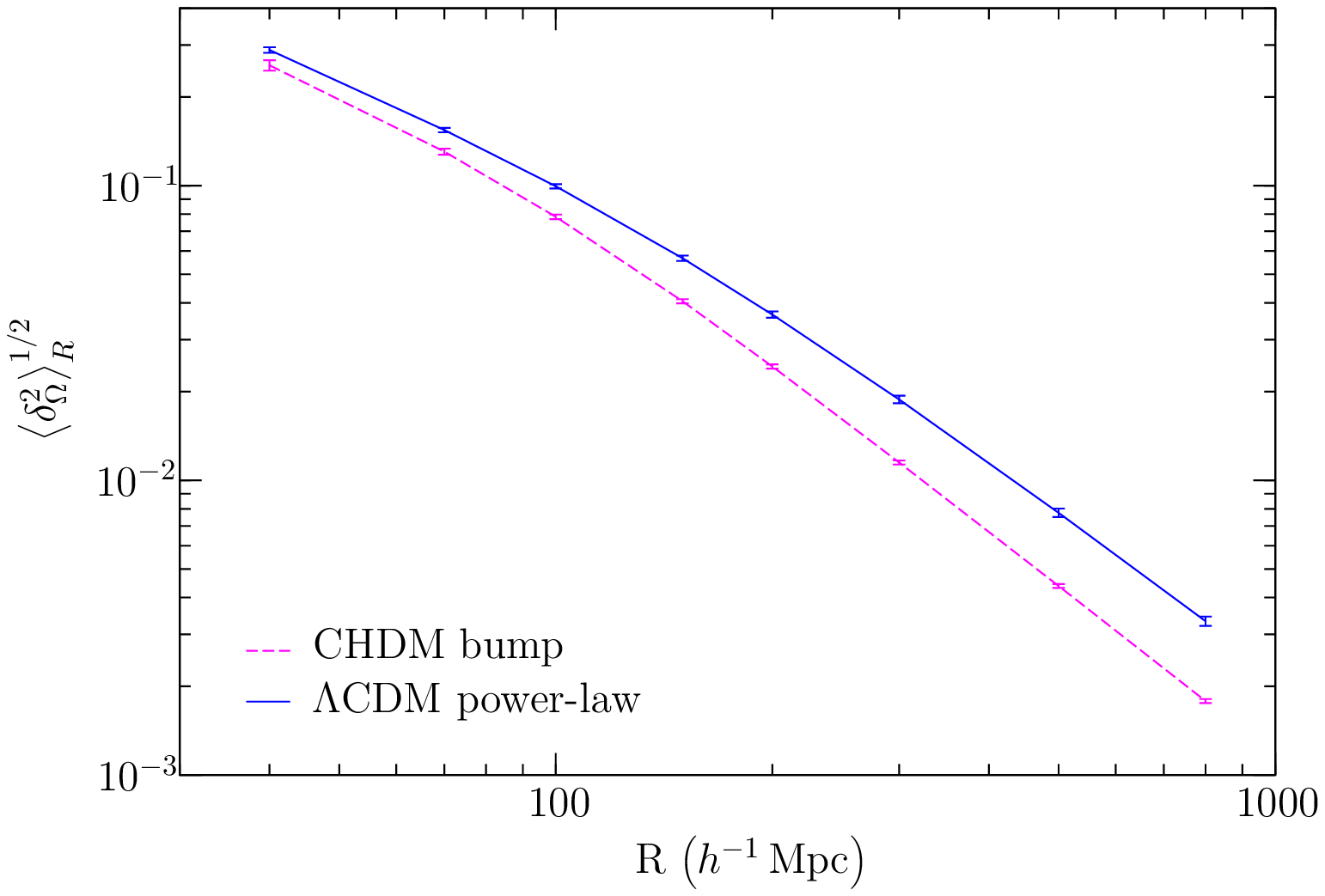} &
\includegraphics*[scale=0.55]{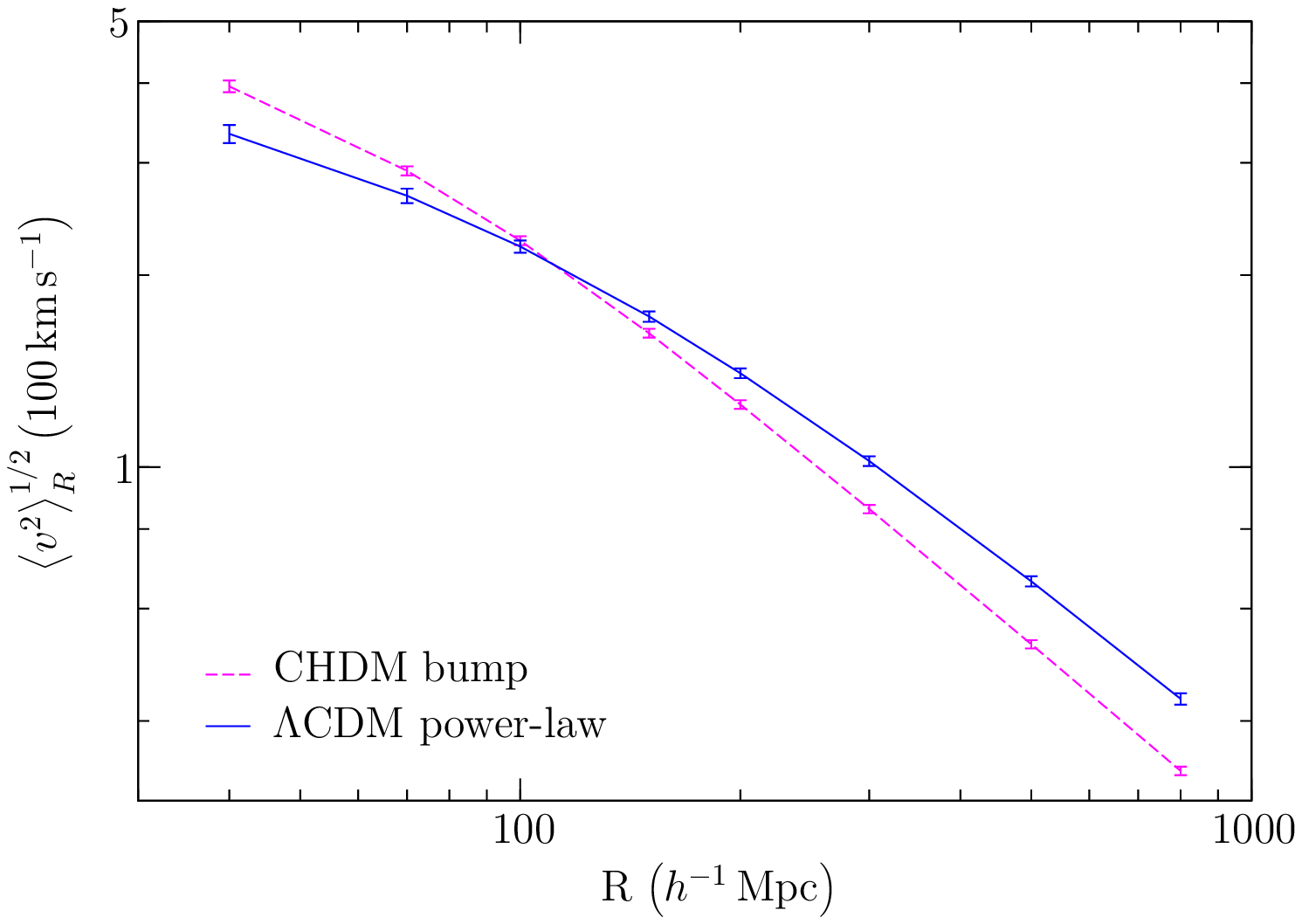}
\end{tabular}
\caption{\label{var} The variation with increasing void radius of the
  variance of the Hubble parameter, the density contrast, the density
  parameter and the peculiar velocity for the $\Lambda$CDM power-law
  and CHDM bump models, given the WMAP-5 and SDSS data (with $1\sigma$
  limits).}
\end{minipage}
\end{figure*}

The mean values of the variances $\left\langle \delta_H^2\right\rangle
_R$, $\left\langle \delta^2\right\rangle _R$, $\left\langle
  {v}^2\right\rangle _R$ and $\left\langle
  \delta_\Omega^2\right\rangle _R$, together with their $1\sigma$
limits, are plotted in Fig.\ref{var}. The different variances in the
two models can be understood with reference to the matter power
spectrum. From the relativistic Poisson equation, a given density
perturbation leads to a larger curvature perturbation in a higher
density universe. Since the amplitude of the primordial curvature
perturbation is similar in both models (as can be seen from
Fig.\ref{wmapfits}) the density contrast during the early matter
dominated era is {\em greater} in the $\Lambda$CDM universe than in
the higher density CHDM universe.  Although the growth of density
perturbations at late times is suppressed in a low density universe,
this means that the matter power spectrum of the $\Lambda$CDM universe
is larger on all scales than that of the CHDM universe, when measured
in units of $h^{-3}\,\mathrm{Mpc}^3$. (This is not evident in
Fig.\ref{wmapfits} where the {\em galaxy} power spectrum is shown ---
the galaxies are more biased in the CHDM universe than in
$\Lambda$CDM so the matter power spectrum is lower.)

This also explains why, as seen in Fig.\ref{var},
$\left\langle\delta^2\right\rangle_R$ is uniformly greater for the
$\Lambda$CDM model. The linear growth factor $f$ is smaller for the
$\Lambda$CDM universe, and the peak in the matter power spectrum
occurs at a larger scale.  Thus the quantity $f^2
\mathcal{P}_\mathrm{m}\left(k\right)$ which appears in eq.(\ref{var1})
is greater for the $\Lambda$CDM universe for wavenumbers below
$k_\mathrm{cross} \simeq 0.01\,h\,\mathrm{Mpc}^{-1}$ but is greater
for the CHDM universe for wavenumbers above $k_\mathrm{cross}$. The
window function $W_H$ (\ref{wh}) makes
$\left\langle\delta_H^2\right\rangle_R$ sensitive to the value of $f^2
\mathcal{P}_\mathrm{m}\left(k\right)$ for the wavenumber
$k\simeq\pi/R$. Consequently the
$\left\langle\delta_H^2\right\rangle_R$ curves for the two models
cross at the scale $\pi/k_\mathrm{cross}\simeq
300$~$h^{-1}\,\mathrm{Mpc}$ as seen in Fig.\ref{var}. The two
$\left\langle {v}^2\right\rangle_R$ curves cross at a smaller scale of
about $100$~$h^{-1}$Mpc. This is because the integral (\ref{var3}) for
the variance in the peculiar velocity $\left\langle
  {v}^2\right\rangle_R$ is more strongly weighted towards small
wavenumbers than the corresponding expression eq.(\ref{var1}) for the
variance in the Hubble contrast
$\left\langle\delta_H^2\right\rangle_R$, which has an additional
factor of $k^2$. Finally for $\left\langle\delta_\Omega^2\right\rangle
_R$ the situation is intermediate between that for the variance in the
density contrast and the variance in the Hubble contrast, since only
some of the terms in $W_\Omega$ contain factors of $f$.

\begin{figure*}
\begin{minipage}{\textwidth}
\centering
\begin{tabular}{cc}
\includegraphics*[angle=0,scale=0.45]{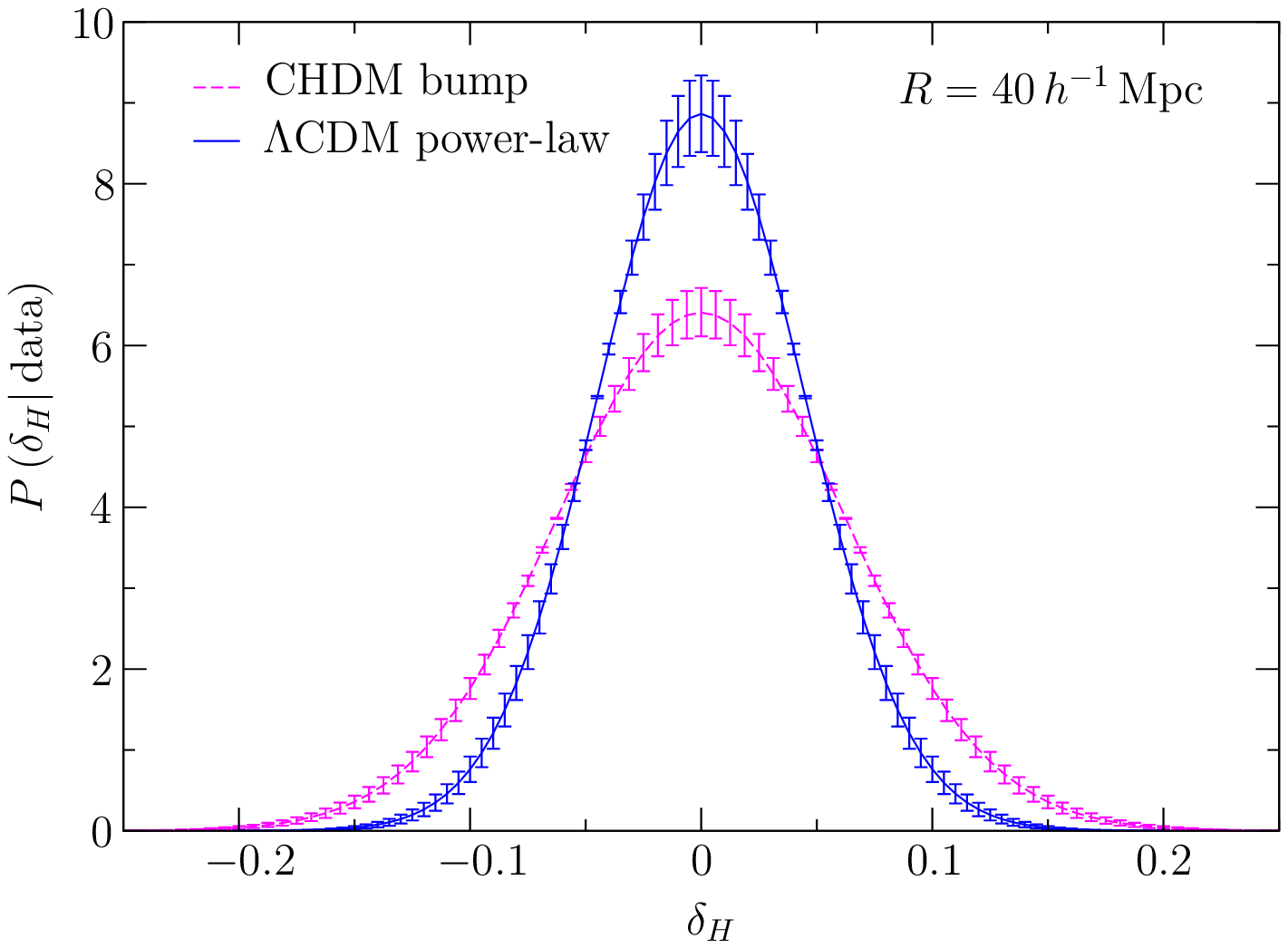} &
\includegraphics*[angle=0,scale=0.45]{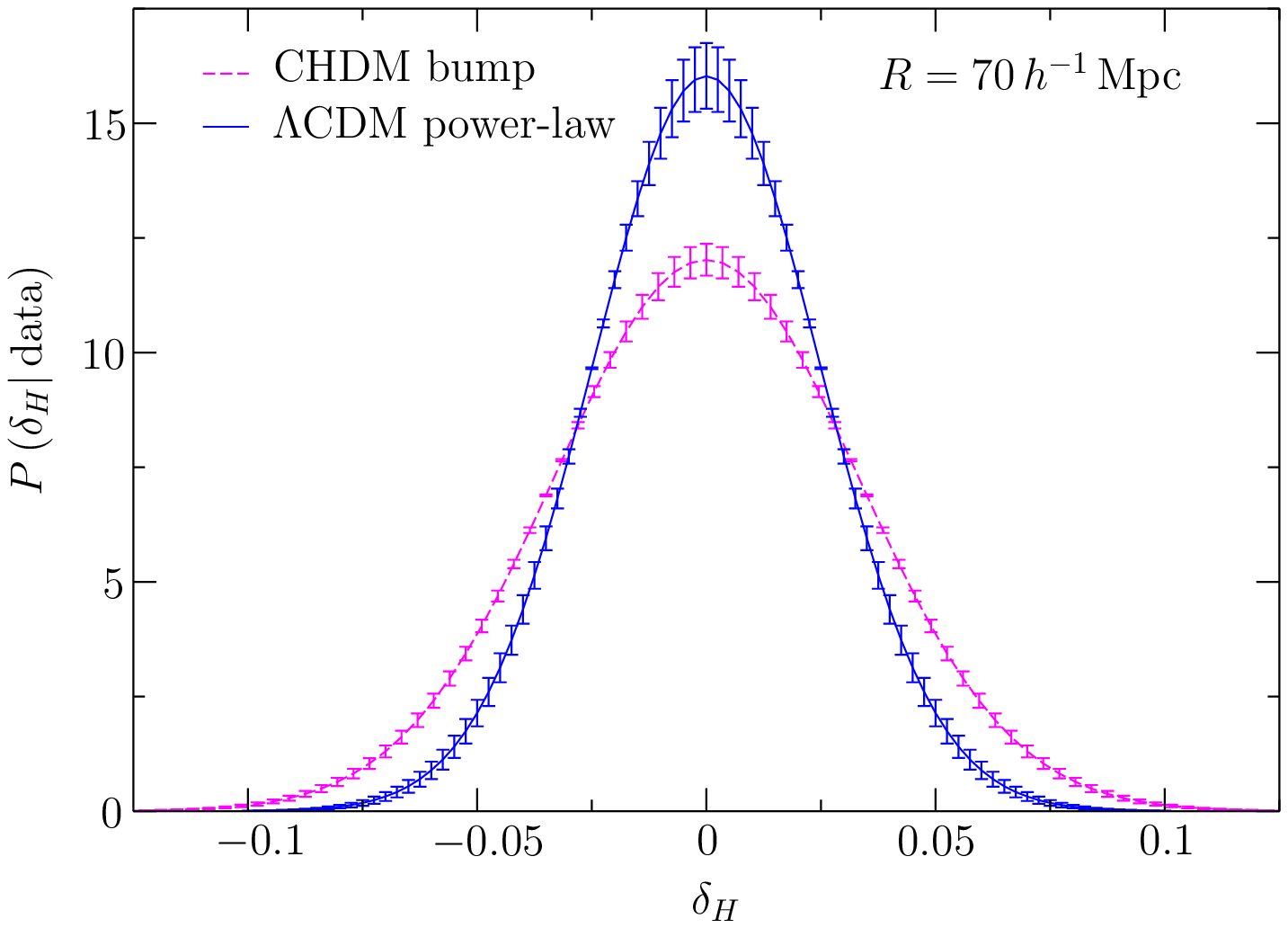} \\
\includegraphics*[angle=0,scale=0.45]{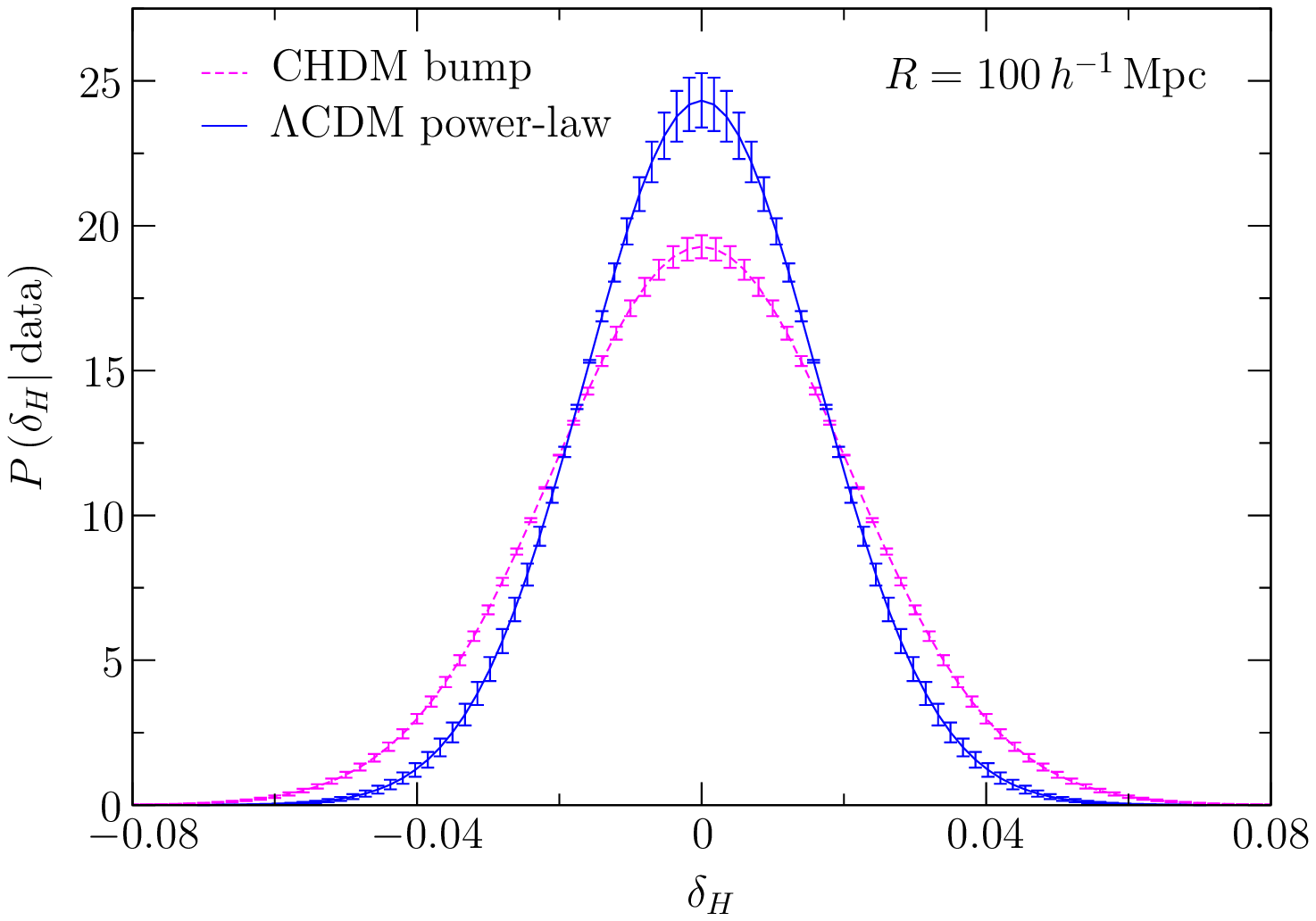} &
\includegraphics*[angle=0,scale=0.45]{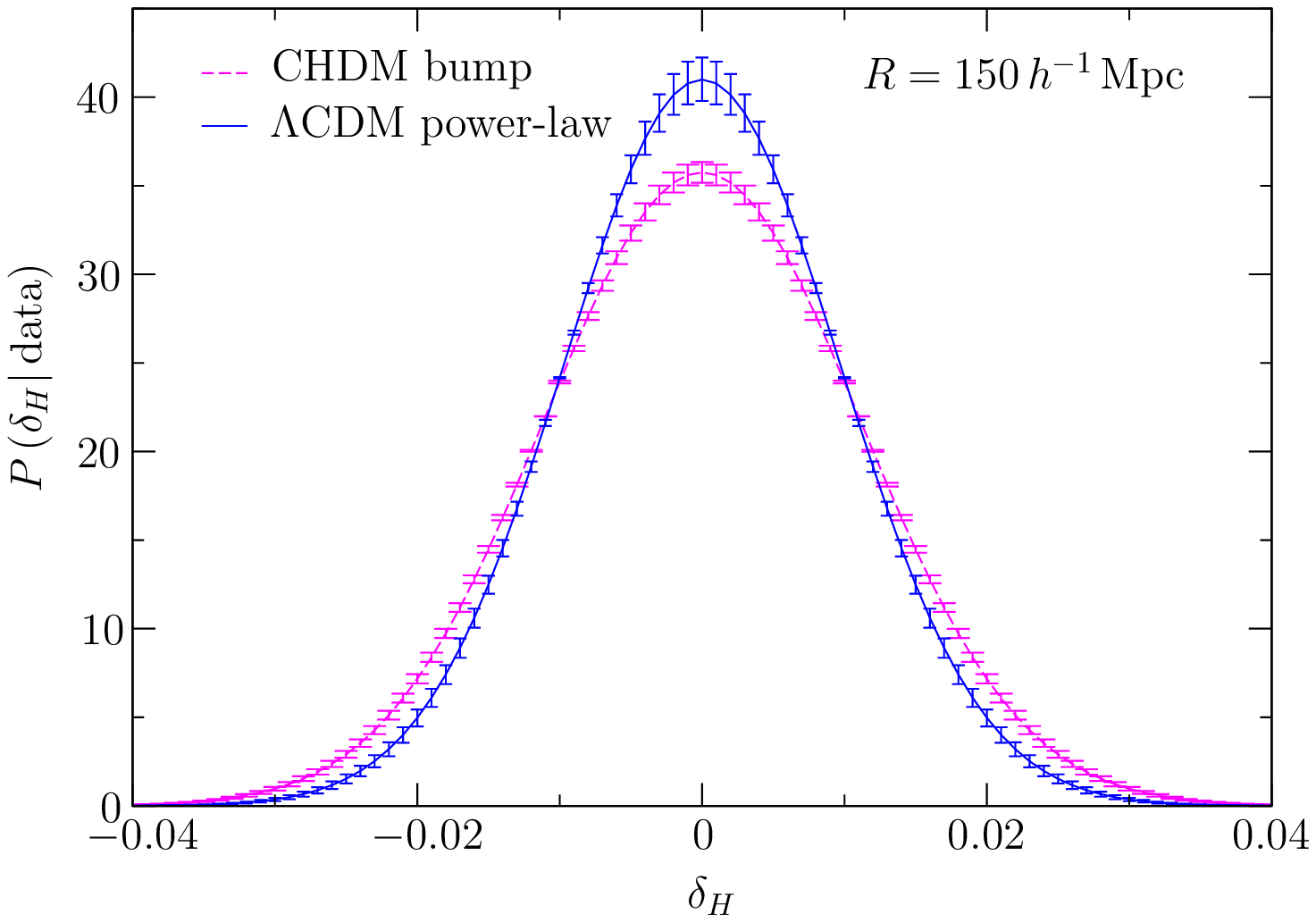} \\
\includegraphics*[angle=0,scale=0.45]{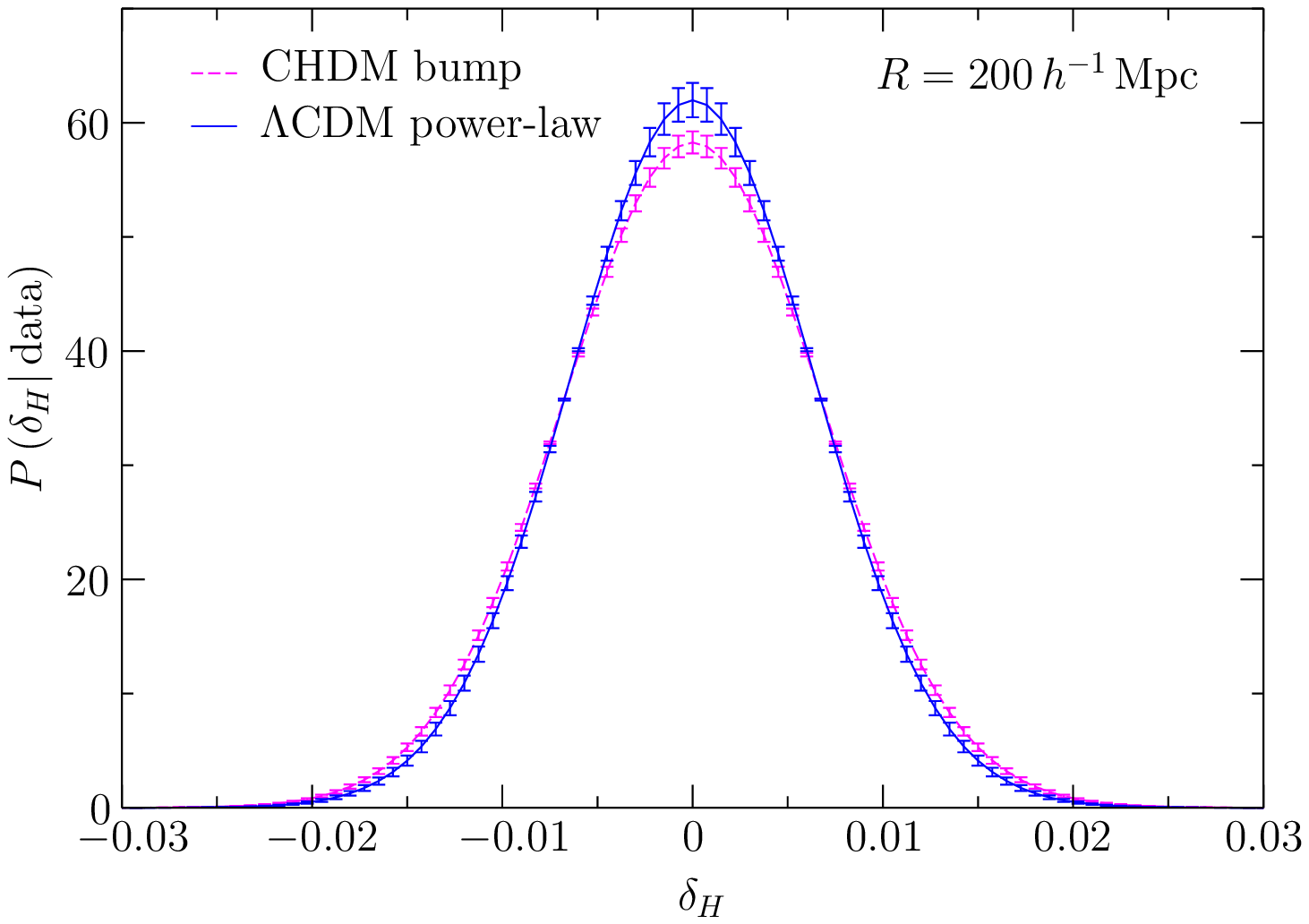} &
\includegraphics*[angle=0,scale=0.45]{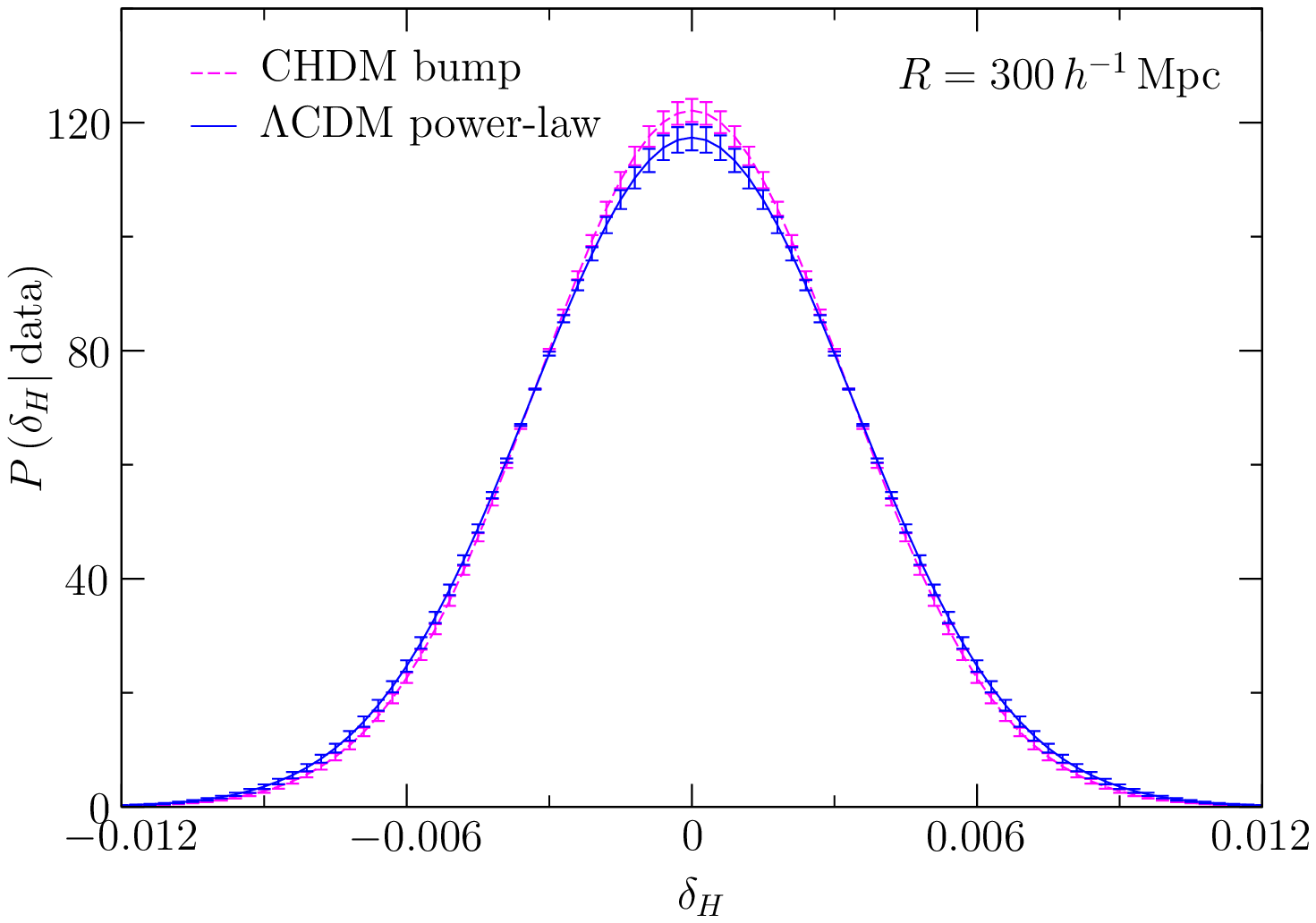} \\
\includegraphics*[angle=0,scale=0.45]{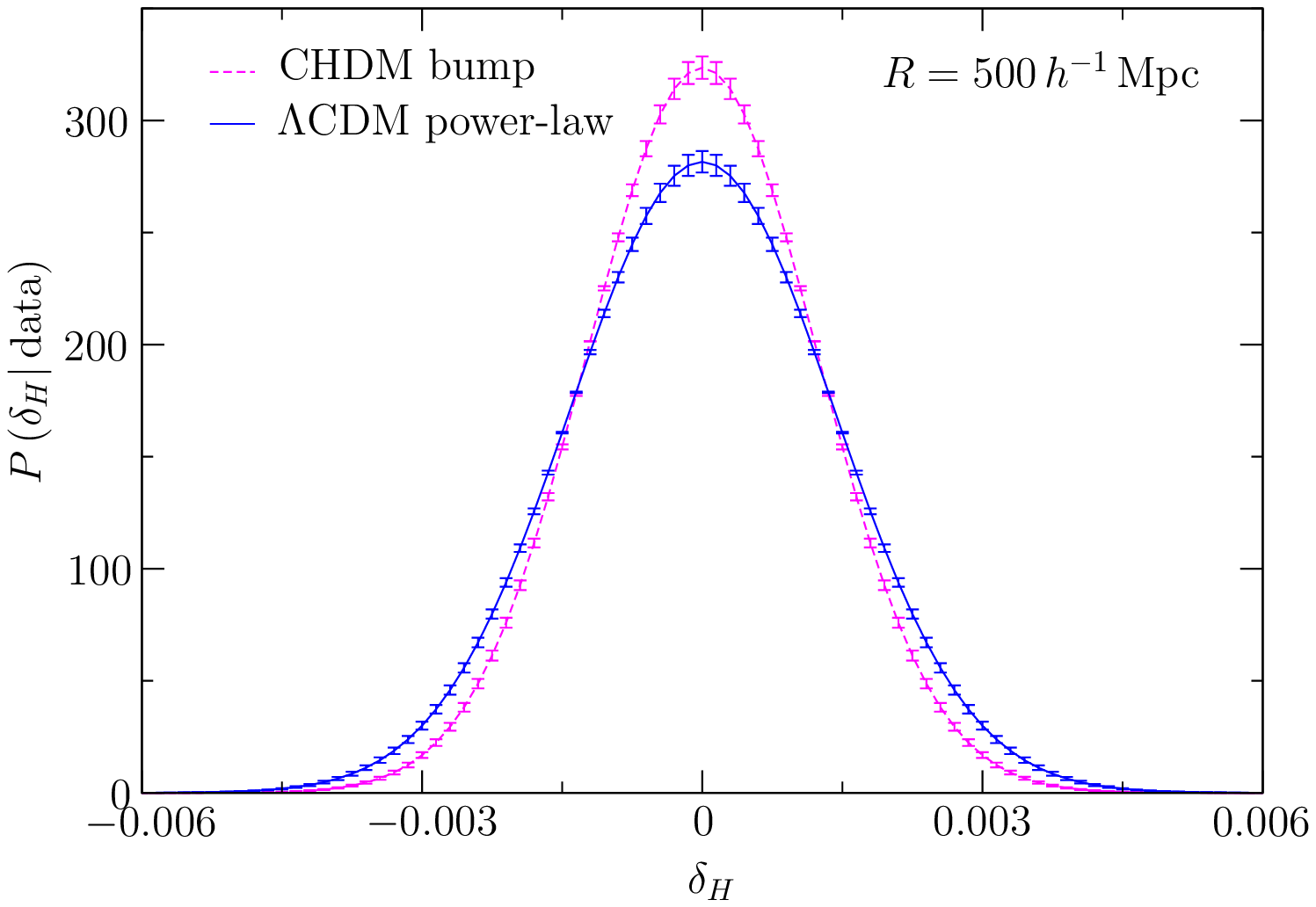} &
\includegraphics*[angle=0,scale=0.45]{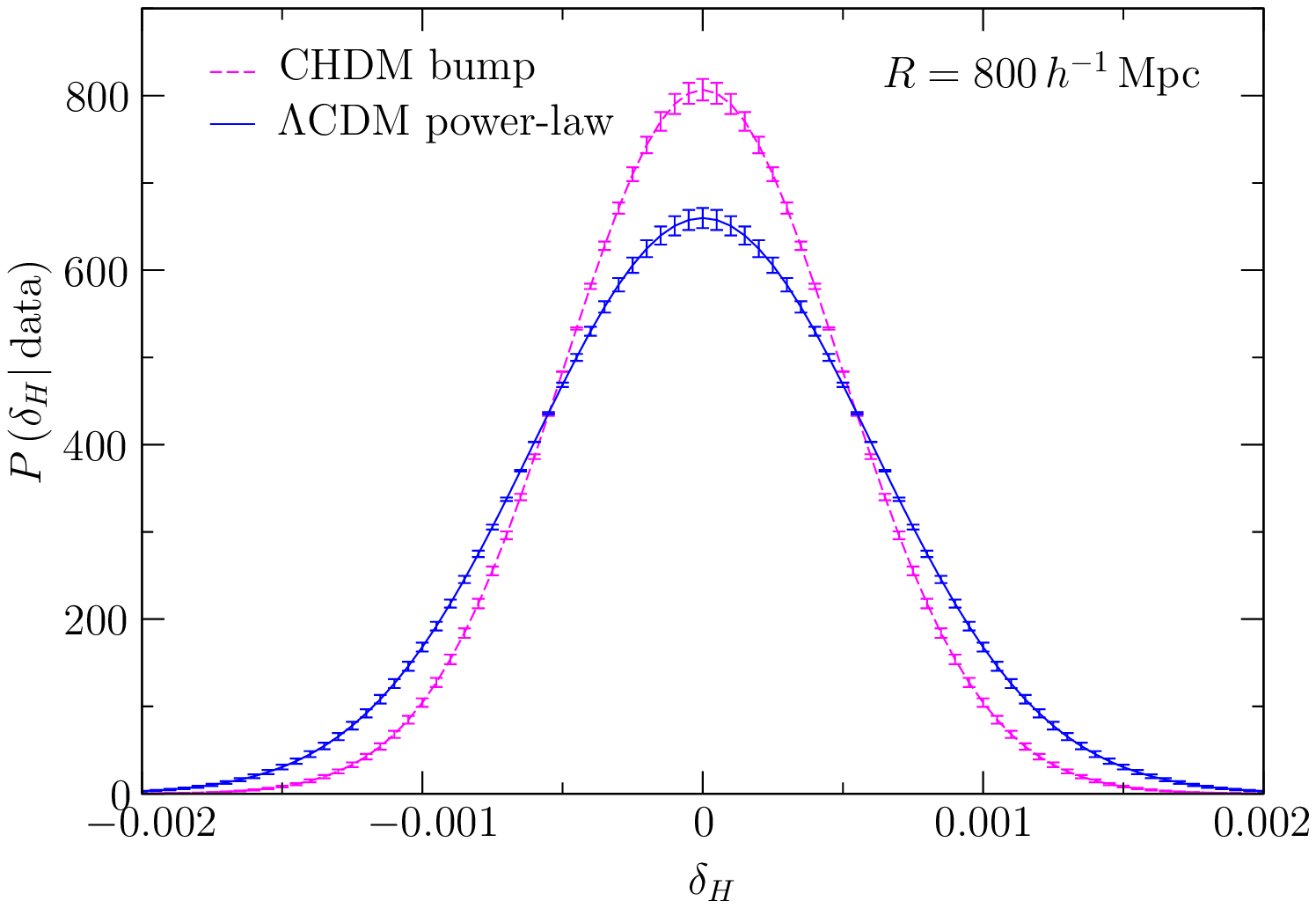}
\end{tabular}
\caption{\label{void} The probability distribution of the Hubble
  contrast (with $1\sigma$ limits), given the WMAP-5 and SDSS data,
  for the $\Lambda$CDM power-law and CHDM bump models, for spherical
  voids of radius $R$ = (40, 70, 100, 150, 200, 300, 500, 800) $\times
  h^{-1}\,\mathrm{Mpc}$.}
\end{minipage}
\end{figure*}

\begin{figure*}
\begin{minipage}{150mm}
\centering
\begin{tabular}{cc}
\includegraphics*[angle=0,scale=0.5]{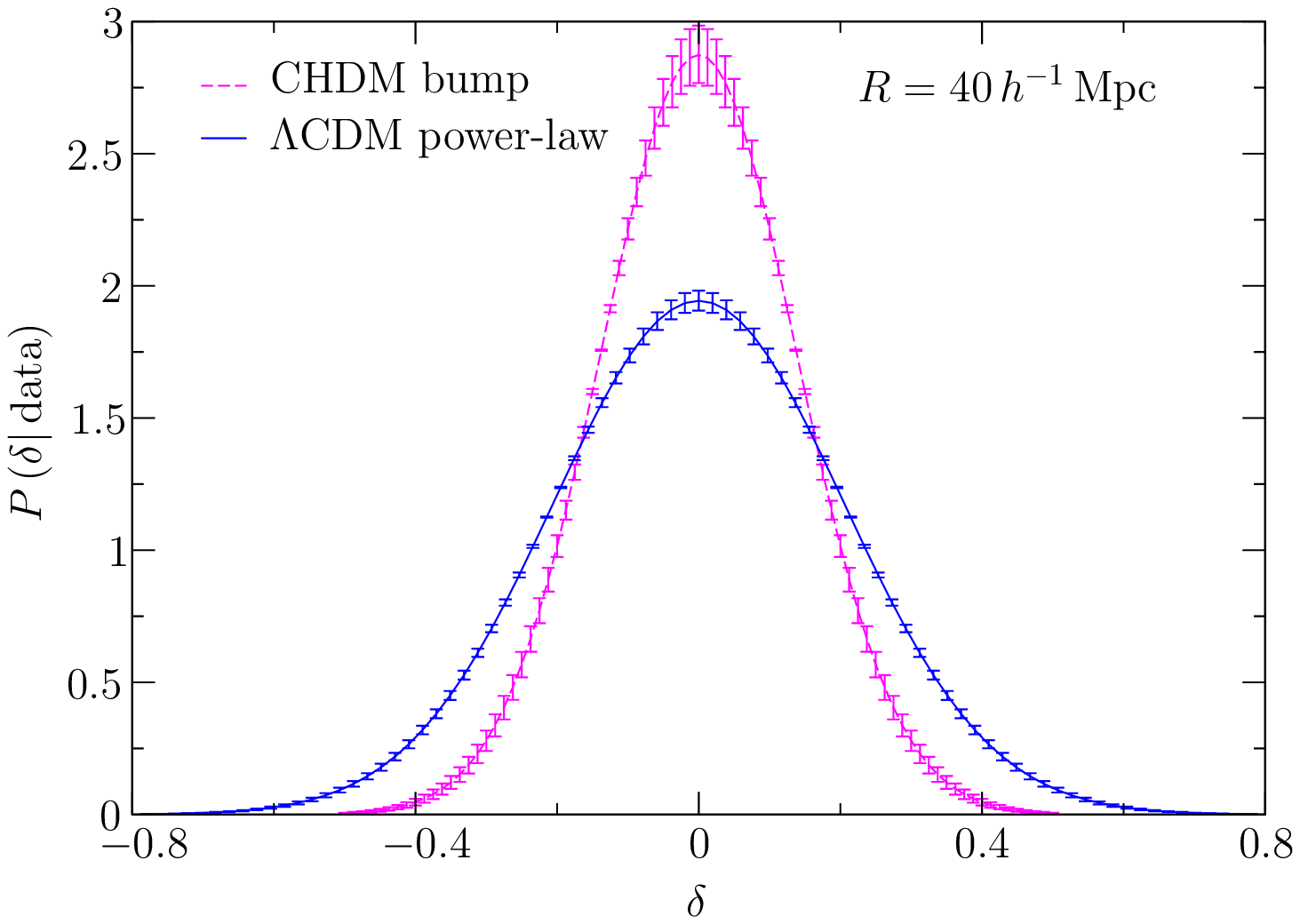} &
\includegraphics*[angle=0,scale=0.5]{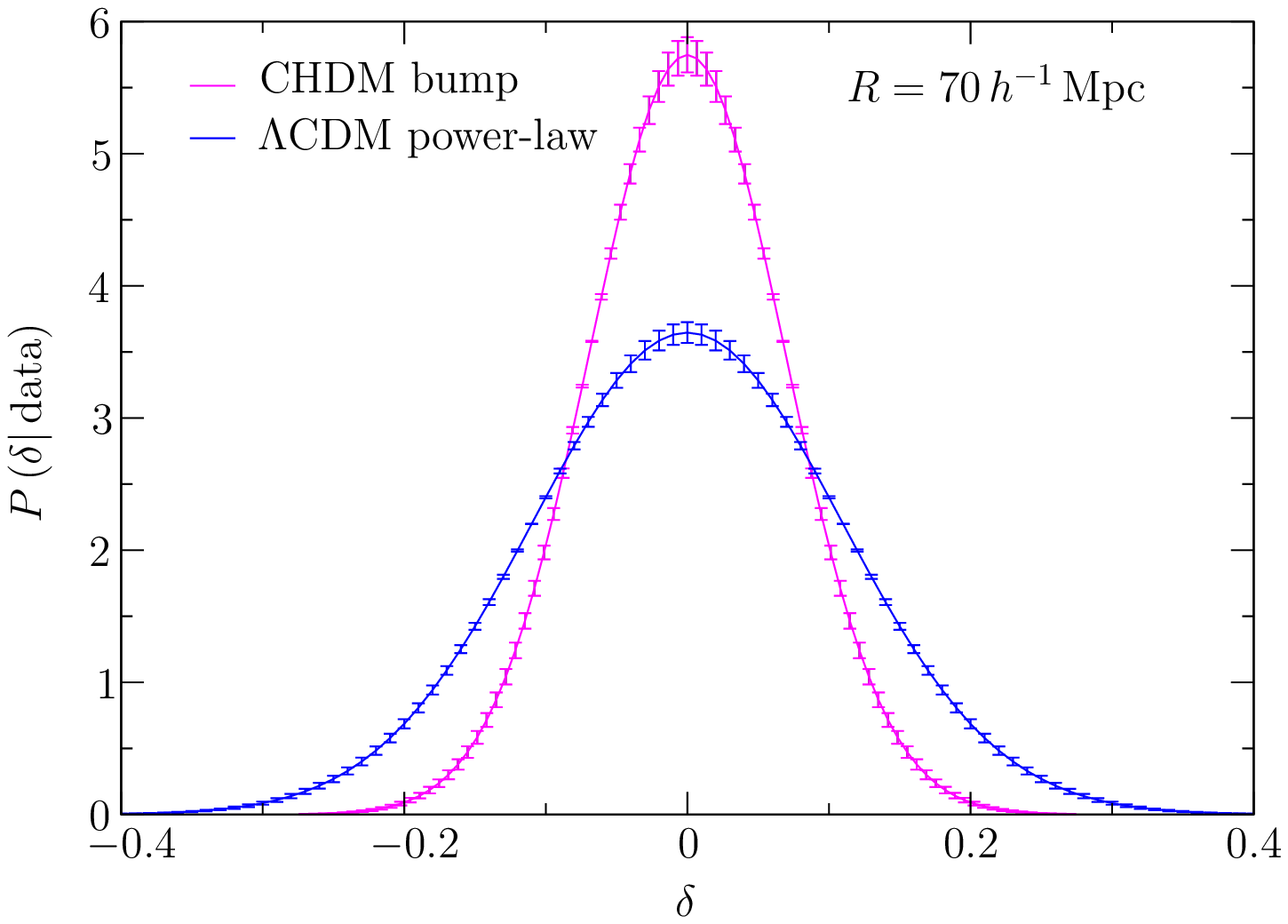} \\ 
\includegraphics*[angle=0,scale=0.5]{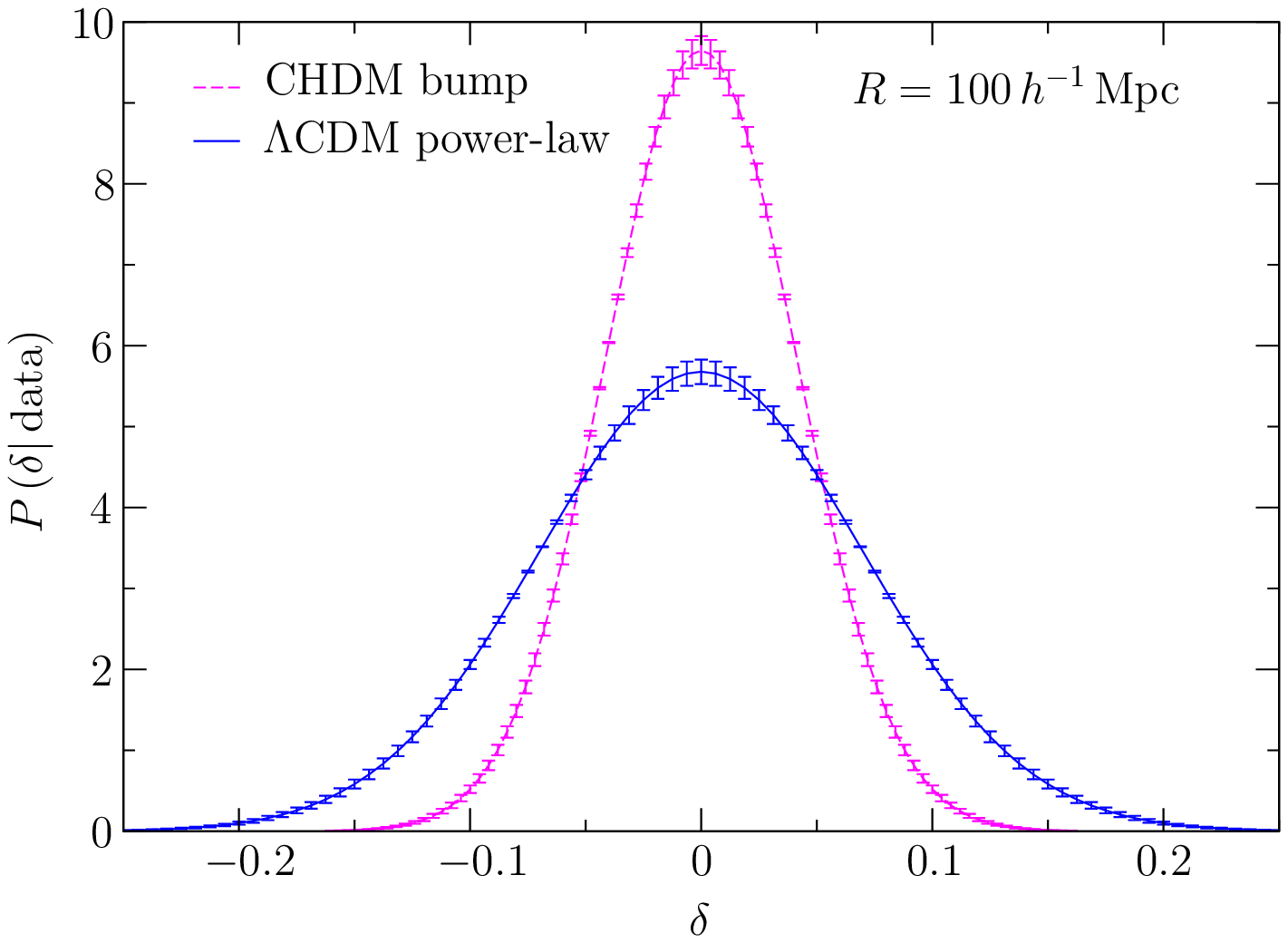} &
\includegraphics*[angle=0,scale=0.5]{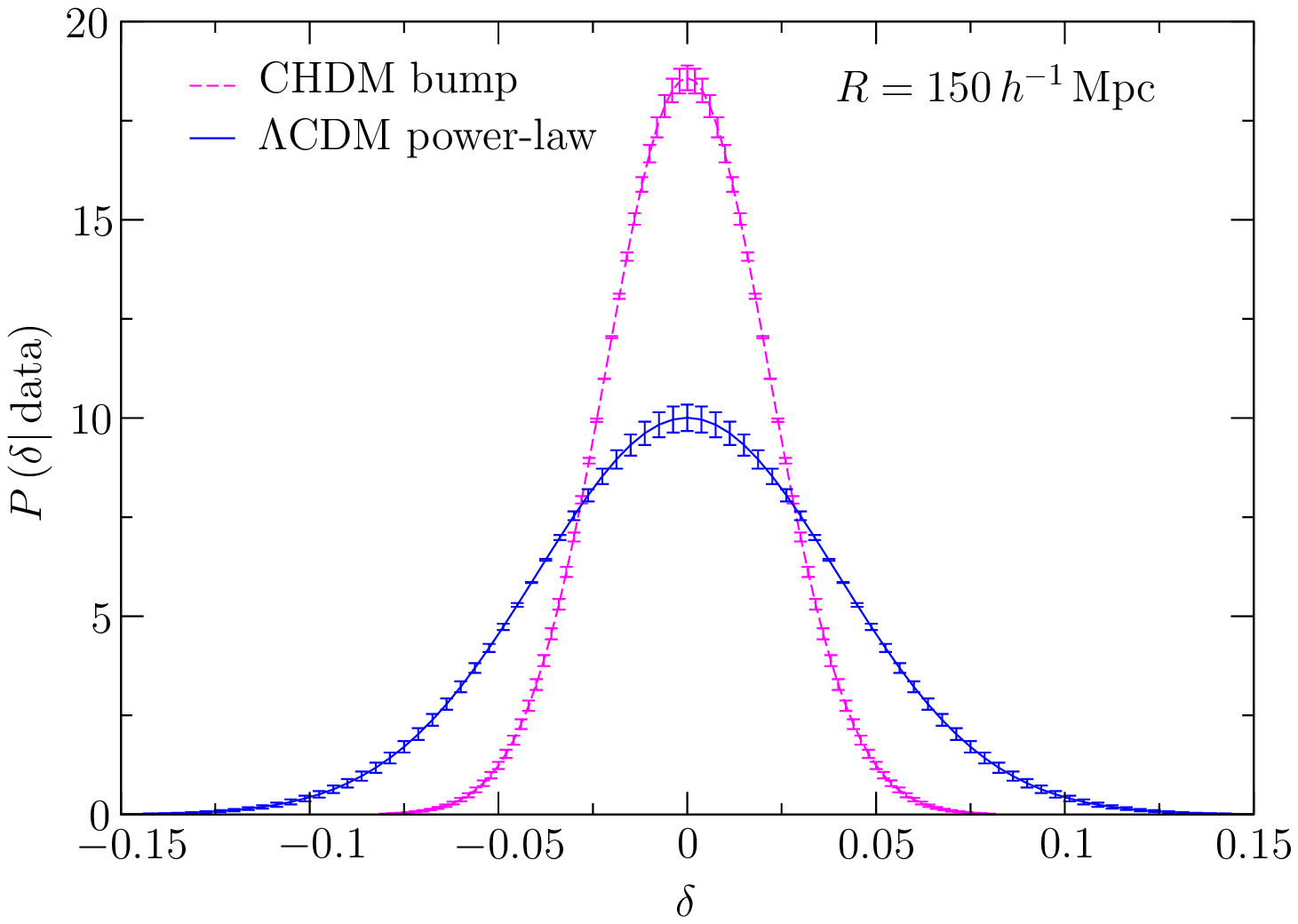} \\
\includegraphics*[angle=0,scale=0.5]{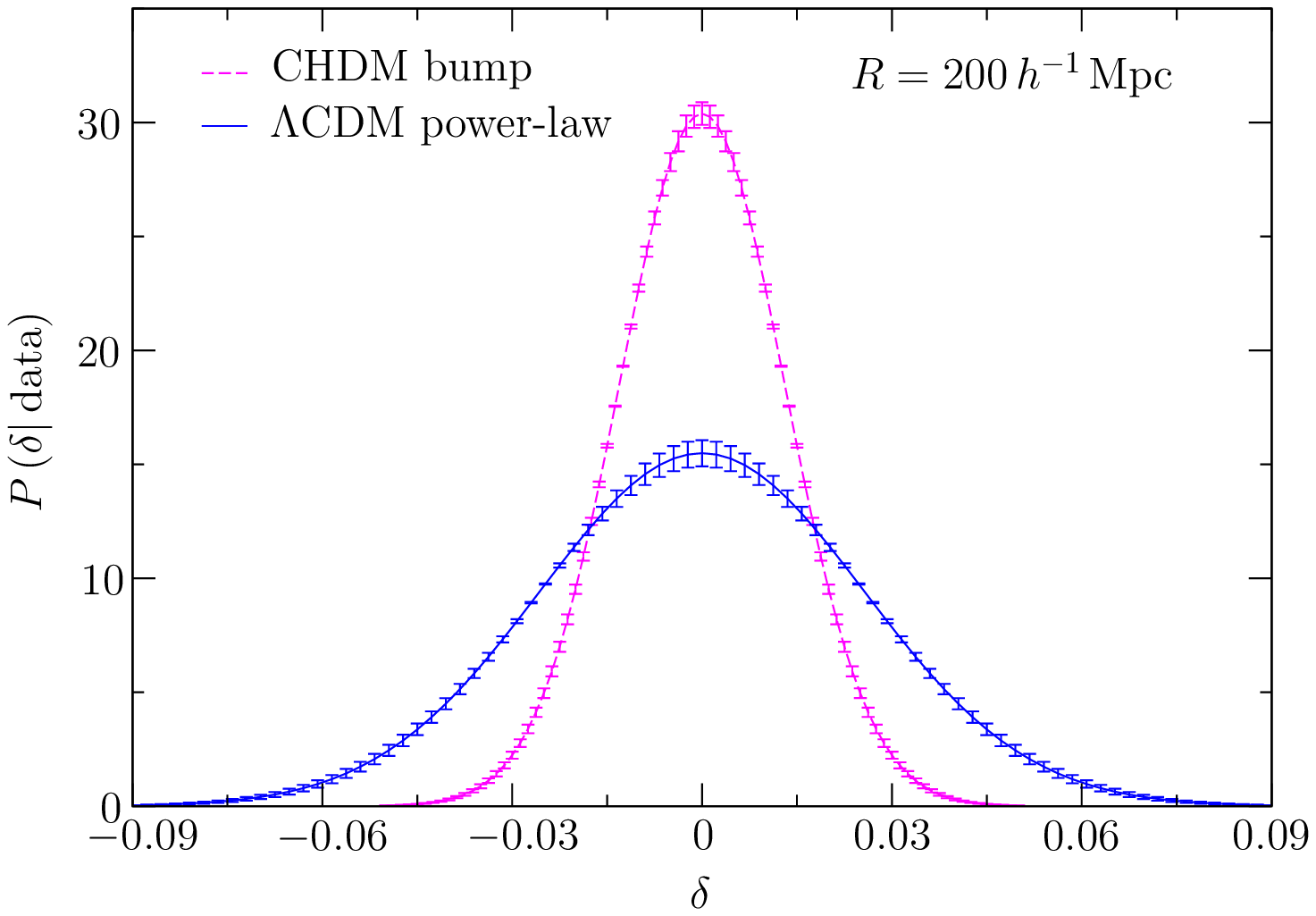} &
\includegraphics*[angle=0,scale=0.5]{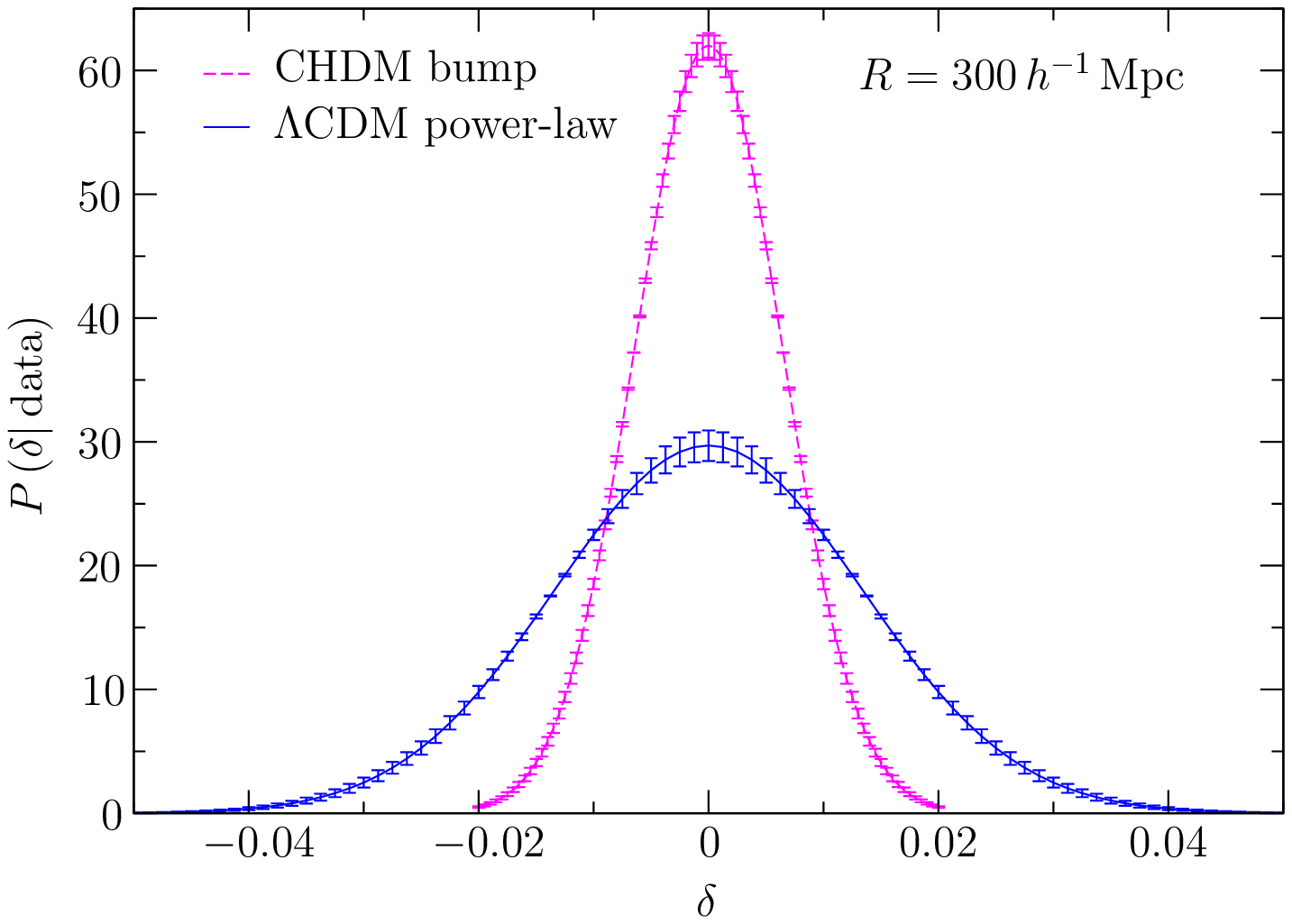} \\
\includegraphics*[angle=0,scale=0.5]{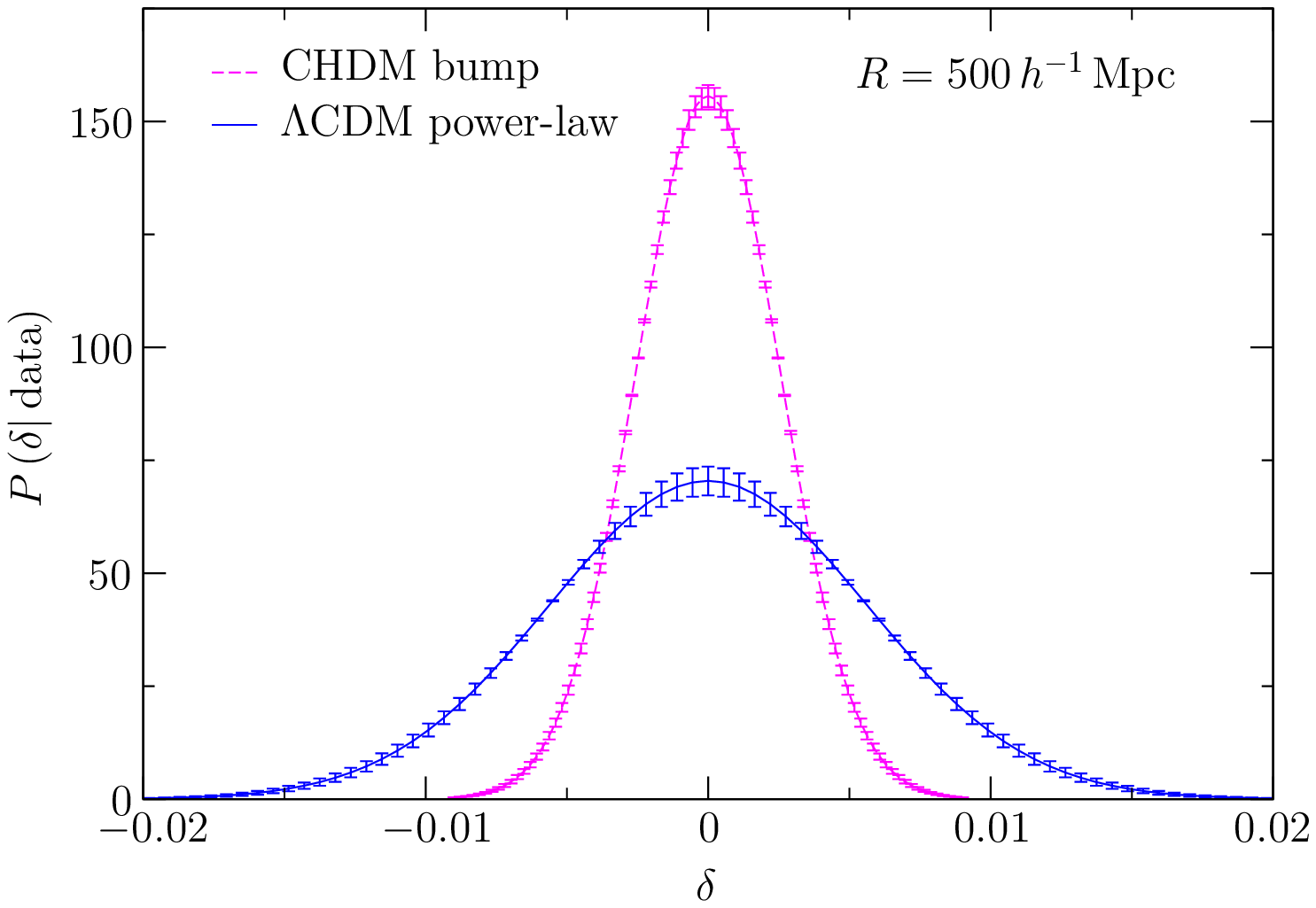} &
\includegraphics*[angle=0,scale=0.5]{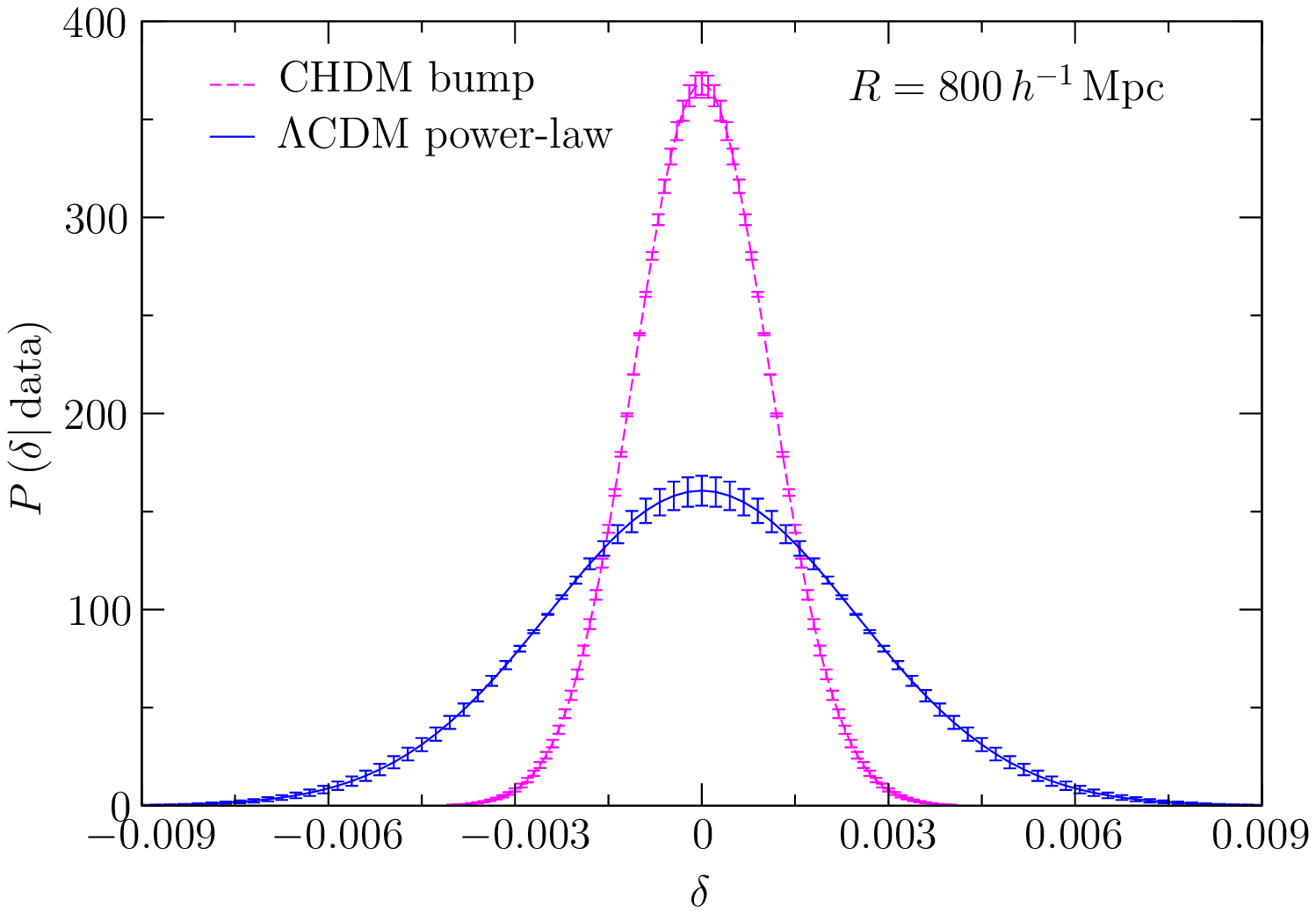}
\end{tabular}
\caption{\label{delta} The probability distribution of the density
  contrast (with $1\sigma$ limits), given the WMAP-5 and SDSS data,
  for the $\Lambda$CDM power-law and CHDM bump models, for spherical
  voids of radius $R$ = (40, 70, 100, 150, 200, 300, 500, 800) $\times
  h^{-1}\,\mathrm{Mpc}$.}
\end{minipage}
\end{figure*}

The scale dependence of $\left\langle\delta_H^2\right\rangle_R$ is the
reason that the $P\left(\delta_{H}\right|\textrm{data})$ distribution
is broader for the $\Lambda$CDM power-law and the CHDM `bump' models
on scales above and below $300$~$h^{-1}\,\mathrm{Mpc}$, respectively,
as shown in Fig.\ref{void}. Similarly the
$P\left(\delta\right|\textrm{data})_R$ distribution is broader for the
$\Lambda$CDM model on all scales, as seen in Fig.\ref{delta}.

\begin{figure*}
\begin{minipage}{\textwidth}
\centering
\begin{tabular}{cc}
\includegraphics*[angle=0,scale=0.5]{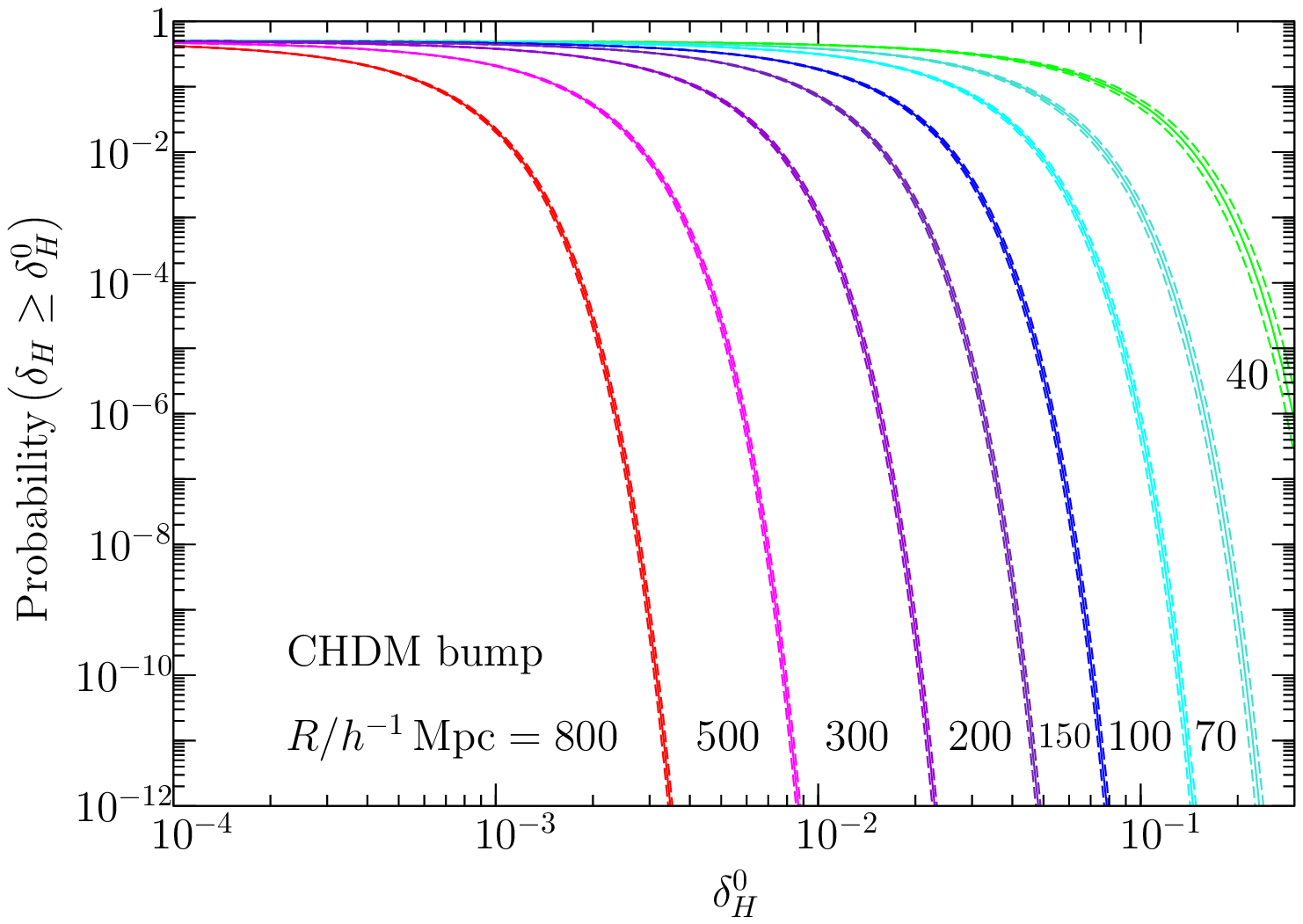} &
\includegraphics*[angle=0,scale=0.5]{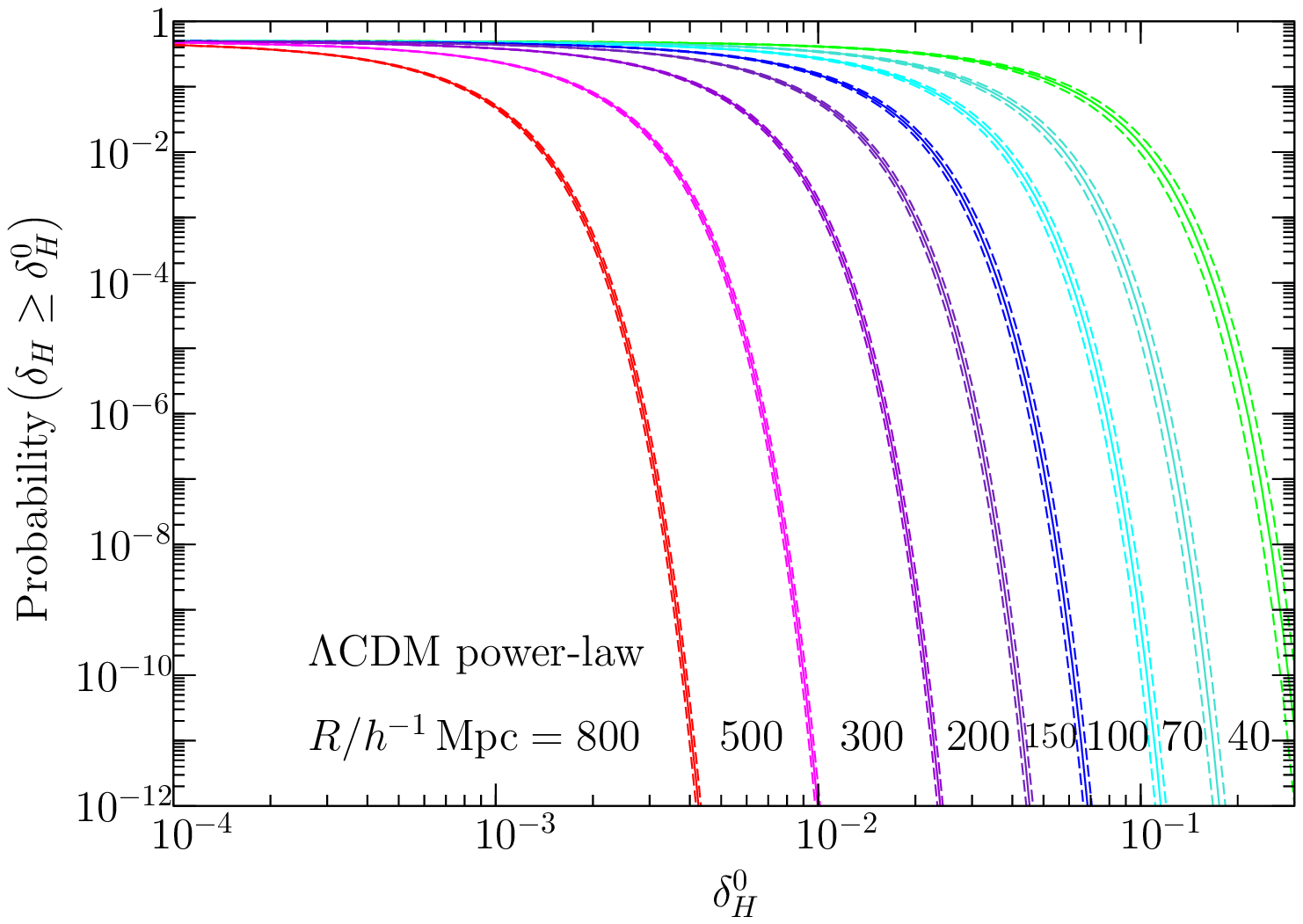}
\end{tabular}
\caption{\label{prob1} The probability of a fluctuation in the Hubble
  contrast greater than or equal to a given value $\delta_H^0$ in a
  sphere of radius $R$ (with $1\sigma$ limits), given the WMAP-5 and
  SDSS data, for the $\Lambda$CDM power-law and CHDM bump models for
  $R$ = (40, 70, 100, 150, 200, 300, 500, 800) $\times
  h^{-1}\,\mathrm{Mpc}$.}
\end{minipage}
\end{figure*}

To illustrate our findings we calculate the probability of a 
fluctuation in the Hubble contrast greater than or equal to a 
given value $\delta_H^0$ in a sphere of radius $R$, given by:
\begin{equation}
  \mathrm{Probability}\left(\delta_{H} \geq 
  \delta_H^0\right)_R=\int_{\delta_H^0}^\infty 
  P\left(\delta_{H}\right|\textrm{data})_R\;\mathrm{d}\delta_H.
\end{equation}
Since $P\left(\delta_{H}\right|\textrm{data})_R$ is symmetric this is
also equivalent to the probability of a fluctuation being less than or
equal to $-\delta_H^0$. As seen in Fig.\ref{prob1}, the probability of
a large excursion in $\delta_H$ is largest on small scales, in
accordance with physical intuition. Note that the probability on all
scales tends to a value of $1/2$ for small $\delta_H^0$ because the
fluctuation has an equal probability of being positive or negative.
The probability is greater for the CHDM model than for the
$\Lambda$CDM model on small scales because the
$P\left(\delta_H{}\right|\textrm{data})_R$ distribution is broader for
the CHDM model on these scales. Conversely since the distribution is
broader on large scales for the $\Lambda$CDM model, the probability is
greater there for this model.

\begin{figure*}
\begin{minipage}{\textwidth}
\centering
\begin{tabular}{cc}
\includegraphics*[angle=0,scale=0.5]{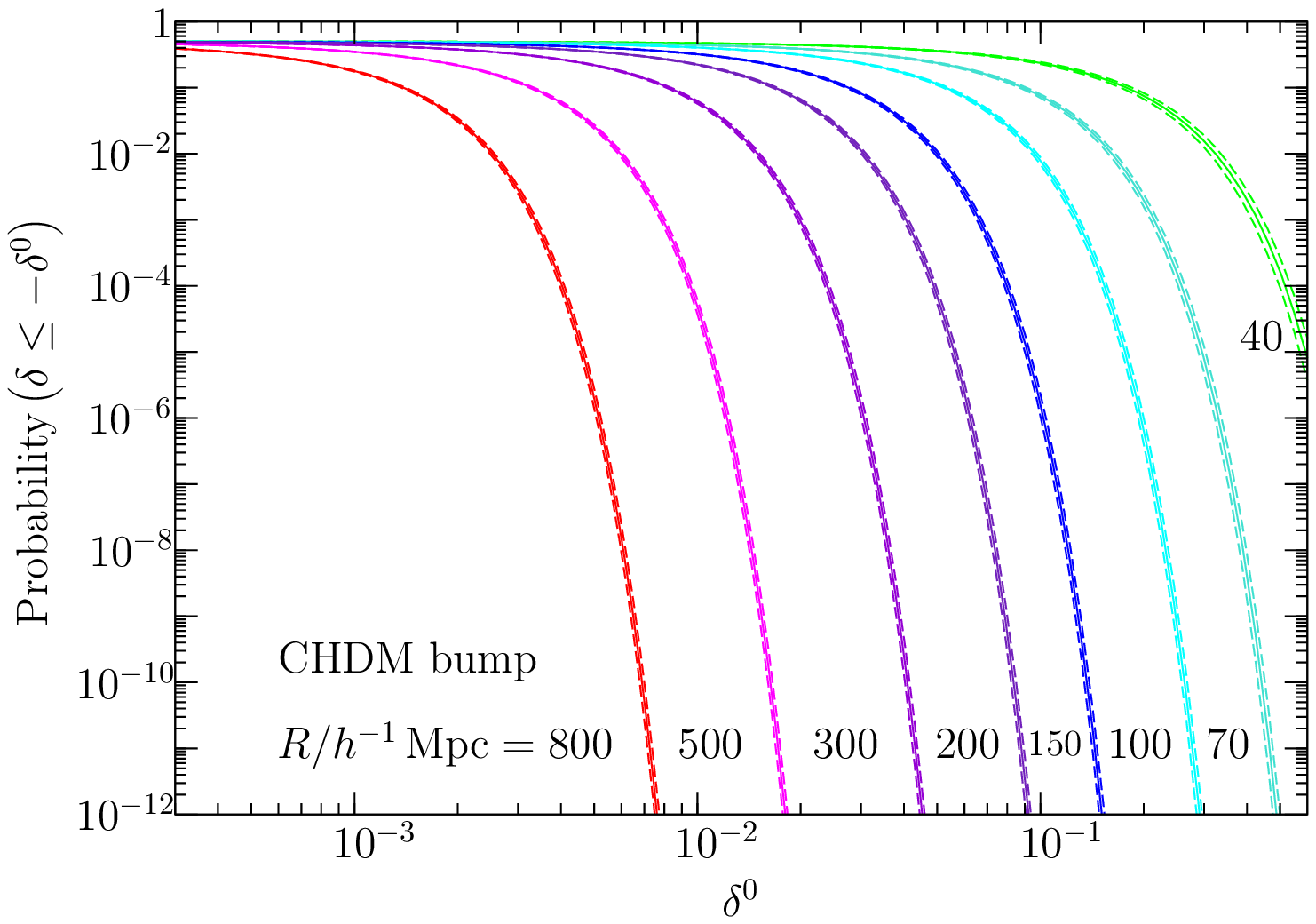} & 
\includegraphics*[angle=0,scale=0.5]{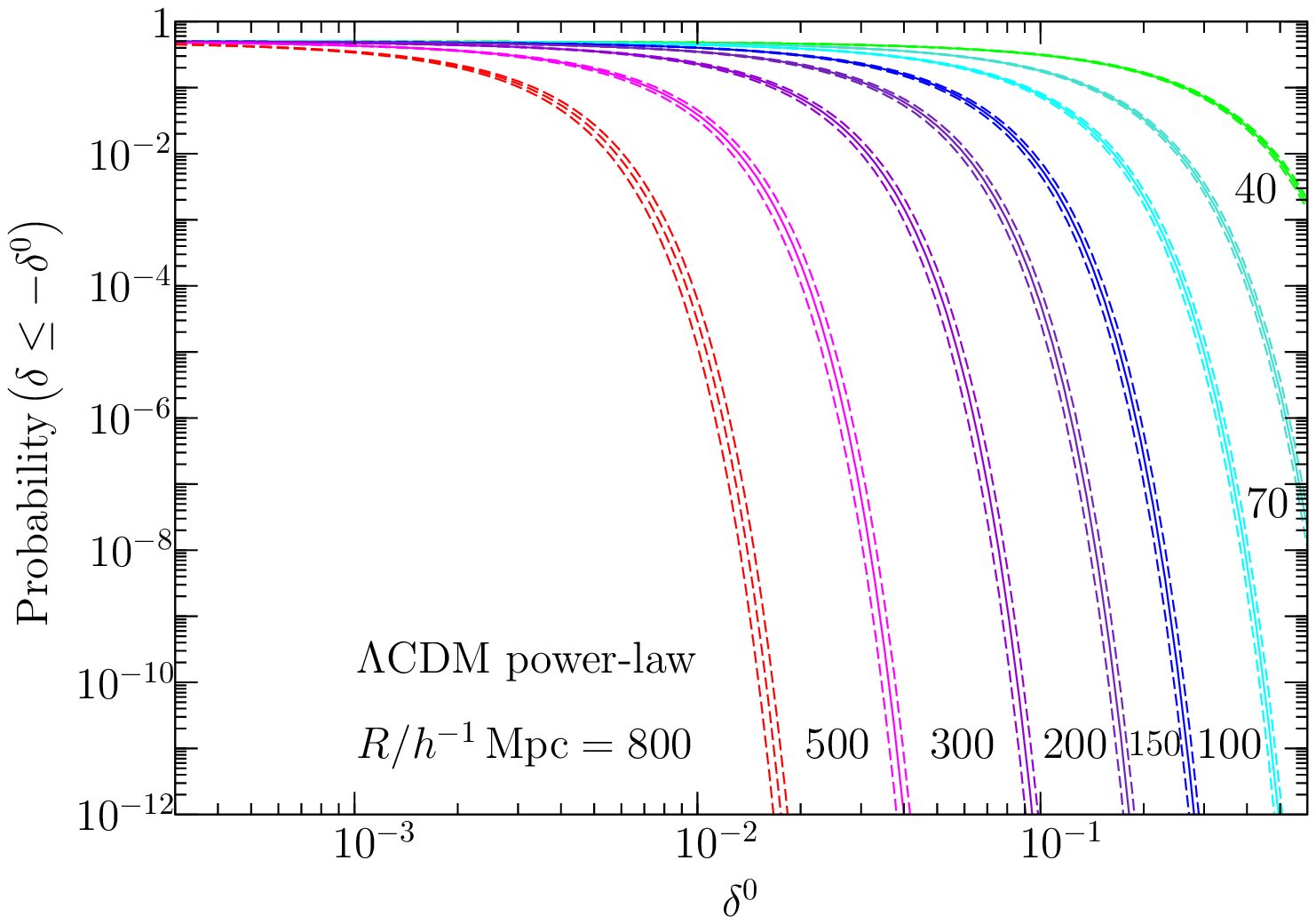}
\end{tabular}
\caption{\label{prob2} The probability of a fluctuation in the density
  contrast less than or equal to a given value $\delta^0$ in a sphere
  of radius $R$ (with $1\sigma$ limits), given the WMAP-5 and SDSS
  data, for the $\Lambda$CDM power-law and CHDM bump models for $R$ =
  (40, 70, 100, 150, 200, 300, 500, 800) $\times
  h^{-1}\,\mathrm{Mpc}$.}
\end{minipage}
\end{figure*}

Similarly we calculate the probability of a fluctuation in the
density contrast less than or equal to a given value $-\delta^0$ in a
sphere of radius $R$, which is given by:
\begin{equation}
\mathrm{Probability}\left(\delta\leq-\delta^0\right)_R=\int^{-\delta^0}_{-\infty} 
P\left(\delta\right|\textrm{data})_R\;\mathrm{d}\delta.
\end{equation} 
This probability is greater for the $\Lambda$CDM model on all scales
as seen in Fig.\ref{prob2}, due to the broader
$P\left(\delta\right|\textrm{data})_R$ distribution.

Moreover, we can determine the probability of one or more voids with
comoving volume $V_1$ occurring within some larger comoving volume
$V_2$. If the ratio $V_2/V_1$ is $N$ to the nearest integer and $p$ is
the probability of a void with volume $V_1$, then the probability of
$n$ voids within $V_2$ is ${N \choose n}p^n\left(1-p\right)^{N-n}$
where ${N \choose n}$ is the binomial coefficient. The expected number
of voids within $V_2$ is $Np$.

\begin{table*}
\begin{minipage}{120mm}
  \caption{\label{tab:parameters} The marginalised cosmological
    parameters for the various models (with 1$\sigma$ limits). The 12
    parameters in the upper section of the Table are varied by
    CosmoMC, while those in the lower section are derived quantities.
    The $\chi^2$ of the fit is given, as is the Akaike information
    criterion relative to the power-law $\Lambda$CDM model.}
\begin{center}
\begin{tabular}{||l|l|l|l||}
\hline
\hline
Parameter & \multicolumn{3}{c||}{Model} \\
\cline{2-4}
& $\Lambda$CDM power-law & CHDM bump with & CHDM bump with \\
&  & $n_1=12$, $n_2=13$ & $n_1$, $n_2$ continuous \\
\hline
\hline
$\Omega_\mathrm{b}h^2$ & $0.02234_{-0.00061}^{+0.00060}$ & $0.01674_{-0.00047}^{+0.00041}$ & $0.01762_{-0.00095}^{+0.00095}$ \\
\hline
$\Omega_\mathrm{c}h^2$ & $0.1144_{-0.0046}^{+0.0046}$ &  &  \\
\hline
$\theta$ & $1.0397_{-0.0031}^{+0.0029}$ & $1.0311_{-0.0039}^{+0.0039}$ & $1.0332_{-0.0047}^{+0.0048}$ \\
\hline
$\tau$ & $0.0842_{-0.0082}^{+0.0077}$ & $0.0721_{-0.0075}^{+0.0069}$ & $0.0771_{-0.0083}^{+0.0073}$  \\
\hline
$f_\nu$ & & $0.114_{-0.012}^{+0.015}$ & $0.085_{-0.022}^{+0.015}$ \\
\hline
$n_\mathrm{s}$ & $0.961_{-0.014}^{+0.014}$ & & \\
\hline
$10^{4}k_{1}/\mathrm{Mpc}^{-1}$ & & $81.7_{-8.3}^{+8.5}$ & $87_{-11}^{+11}$ \\
\hline
$10^{4}k_{2}/\mathrm{Mpc}^{-1}$ & & $442_{-53}^{+47}$ & $500_{-54}^{+21}$  \\
\hline
$\ln\left(10^{10}\mathcal{P_R}\right)$ & $3.078_{-0.037}^{+0.037}$ &  &  \\
\hline
$\ln\left(10^{10}\mathcal{P_R}^{(0)}\right)$ & & $3.294_{-0.031}^{+0.031}$ & $3.274_{-0.048}^{+0.048}$ \\
\hline
$\ln\left(10^{10}\mathcal{P_R}^{(1)}\right)$ & & & $3.462_{-0.036}^{+0.036}$ \\
\hline
$\ln\left(10^{10}\mathcal{P_R}^{(2)}\right)$ & & & $3.183_{-0.041}^{+0.043}$ \\
\hline
\hline
$\Omega_\mathrm{c}h^2$  & & $0.1450_{-0.0077}^{+0.0079}$ & $0.156_{-0.013}^{+0.012}$ \\
\hline
$\Omega_\mathrm{d}h^2$ & & $0.1634_{-0.0045}^{+0.0042}$ & $0.1702_{-0.0074}^{+0.0073}$ \\
\hline
$h$ & $0.695_{-0.021}^{+0.021}$ & $0.4244_{-0.0055}^{+0.0052}$ & $0.4333_{-0.0094}^{+0.0093}$ \\
\hline
Age/Gyr & $13.78_{-0.14}^{+0.14}$ & $15.36_{-0.19}^{+0.20}$ & $15.05_{-0.32}^{+0.33}$  \\
\hline
$\Omega_\mathrm{m}$ & $0.284_{-0.025}^{+0.025}$ & &  \\
\hline
$\Omega_\mathrm{\Lambda}$ & $0.716_{-0.025}^{+0.025}$ & &  \\
\hline
$\sigma_8$ & $0.817_{-0.027}^{+0.027}$ & $0.617_{-0.055}^{+0.059}$ &  $0.700_{-0.098}^{+0.098}$ \\
\hline
$z_\mathrm{reion}$ & $11.0_{-1.4}^{+1.4}$ & $13.0_{-2.0}^{+2.0}$ & $13.4_{-2.0}^{+2.1}$ \\
\hline
$\Delta m_1^2$ & & $0.07495_{-0.00046}^{+0.00046}$ & $0.089_{-0.020}^{+0.020}$ \\
\hline
$\Delta m_2^2$ & &  $0.15133_{-0.00084}^{+0.00084}$ & $0.136_{-0.016}^{+0.015}$ \\
\hline
$H\left(t_2-t_1\right)$ & &  $1.68_{-0.13}^{+0.12}$ & $1.73_{-0.15}^{+0.14}$ \\
\hline
\hline
$\chi^2$ & 1339.9 & 1339.9 & 1330.2 \\
\hline
$\Delta_\mathrm{AIC}$ & $0$ & $6.0$ & $-3.7$ \\
\hline
\hline
\end{tabular}
\end{center}
\end{minipage}
\end{table*} 

\section{Discussion}

A void with $\delta_H\simeq 0.2-0.3$ and a radius exceeding
$100~h^{-1}\,\mathrm{Mpc}$ is required to fit the supernova data
without dark energy
\citep{Tomita:2001gh,Biswas:2006ub,Alexander:2007xx}. The probability
that we are situated in such a void is less than $10^{-12}$ as can be
seen from Fig.\ref{prob1}. The probability is exponentially smaller for
the larger voids of Gpc size that have also been considered
\citep{Alnes:2005rw,GarciaBellido:2008nz,Clifton:2008hv}.\footnote{
  There is a further constraint on Gpc scale voids from the observed
  absence of a `$y$-distortion' in the spectrum of the CMB
  \citep{Caldwell:2007yu} and from the `kinetic Sunyaev-Zeldovich'
  effect observed for X-ray emitting galaxy clusters
  \citep{GarciaBellido:2008gd}. However this has no impact on smaller
  voids.}

However before we dismiss the possibility of a local void on these
grounds we should also evaluate the probability of voids which have
actually been claimed to exist elsewhere in the universe. For example
it has been argued that a void with radius
$200-300~h^{-1}\,\mathrm{Mpc}$ and an density contrast of
$\delta=-0.3$ at $z\sim1$ can account for the WMAP `cold spot' in a
$\Lambda$CDM universe \citep{Inoue:2006rd}. Even if we 
conservatively take the radius to be
$150~h^{-1}\,\mathrm{Mpc}$ (and the same underdensity), the probability
that one or more such voids lie within the volume out to $z=1$ is only
$1.05_{-0.93}^{+5.24}\times10^{-10}$.

It has been argued that the WMAP cold spot may not be a localized
feature \citep{Naselsky:2007yd} and there may be no matching void in
the NVSS radio source catalogue \citep{Smith:2008tc}, however an
equally striking anomaly arises if we consider the large number of
voids which have been identified in the SDSS LRG survey in a search
for the late ISW effect \citep{Granett:2008xb,Granett:2008ju}. These
are of angular radius $\sim 4^0$ corresponding to a (comoving) radius
of $\sim 50~h^{-1}\,\mathrm{Mpc}$ and are tabulated as having
$1\sigma$, $2\sigma$ or $3\sigma$ underdensities. These numbers relate
to the detection significance (the likelihood of detecting the void by
chance out of a Poisson distribution) rather than the likelihood of
finding such underdensities in a gaussian field which we have computed
in this paper (B. Granett, private communication). Moreover the
observed LRGs are biased with regard to the dark matter hence the
underdensities in dark matter are likely to be smaller than the quoted
values.

\begin{figure*}
\begin{minipage}{\textwidth}
\centering
\begin{tabular}{cc}
\includegraphics*[angle=0,scale=0.5]{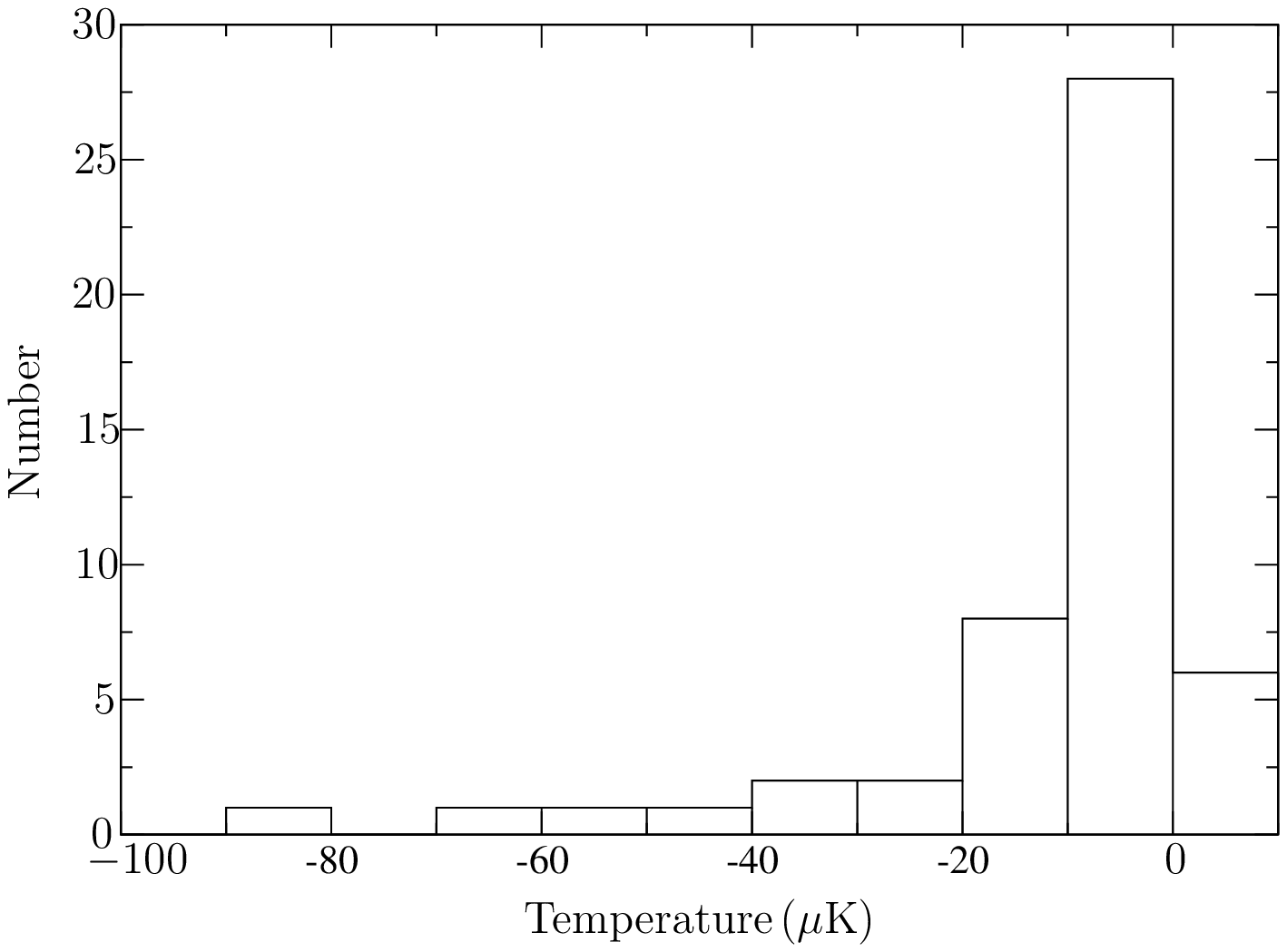} &
\includegraphics*[angle=0,scale=0.5]{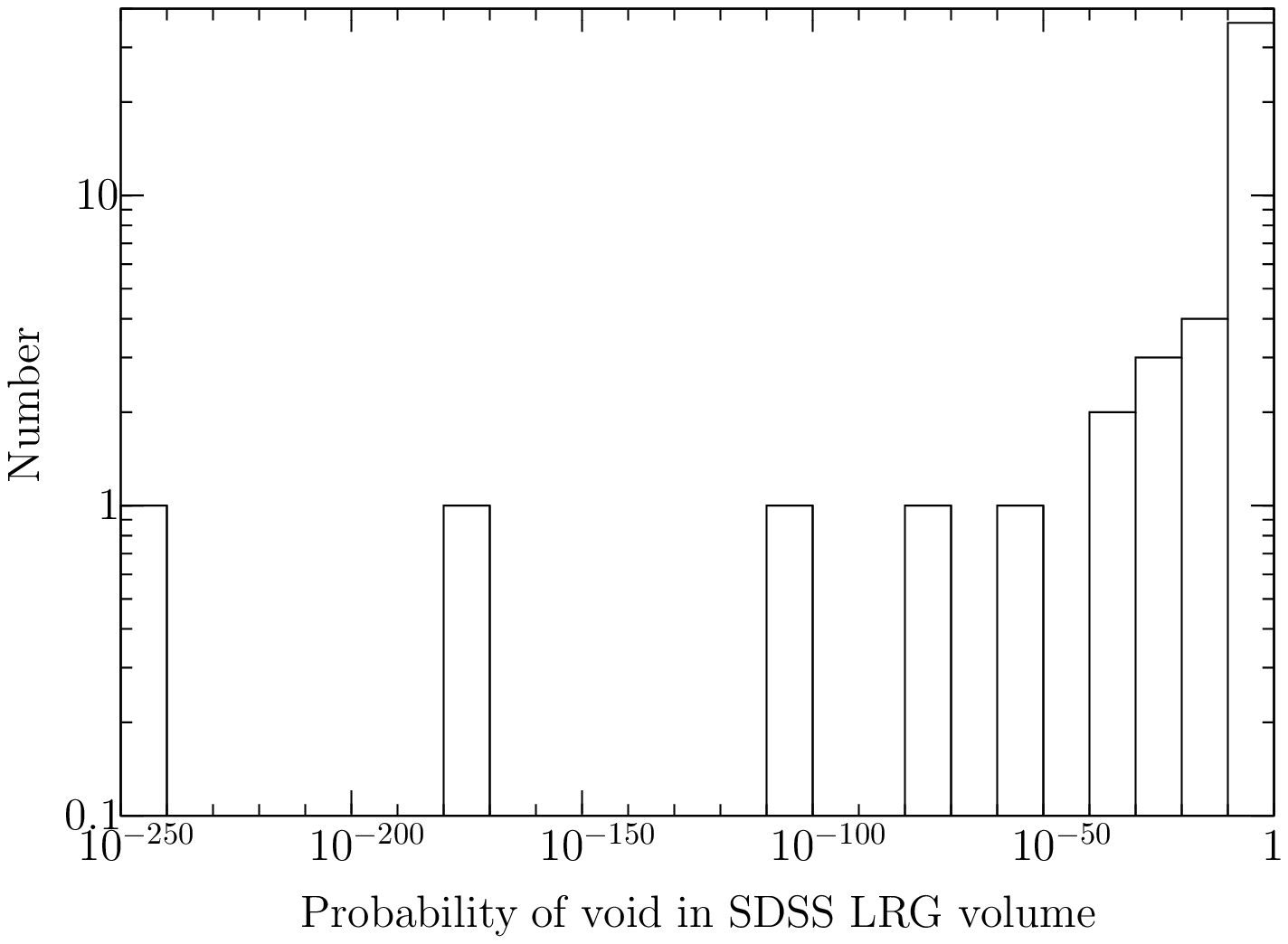}
\end{tabular}
\caption{\label{hist} The left panel shows the ISW signals of the 50
  voids detected by \citet{Granett:2008xb}, calculated using
  eq.(\ref{isw}); in order to match the observed average ISW signal of
  $-11.3 \,\mu \mathrm{K}$ it has been necessary to increase the void
  radii by a factor of 1.75 and the underdensities by a factor of
  5. The right panel show the probability of such voids occurring in
  the SDSS LRG survey volume according to the concordance $\Lambda$CDM
  model.}
\end{minipage}
\end{figure*}

However if \citet{Granett:2008xb,Granett:2008ju} have indeed detected
the late ISW effect as they assert, we can simply circumvent these
uncertainties by requiring that the voids be large enough and/or
underdense enough to yield the {\em observed} CMB temperature
decrements. To calculate the late ISW effect we consider the
propagation of CMB photons to us from the last scattering surface
through an intervening void. The photon temperature change caused by
the void is
\begin{equation}
\frac{\Delta T}{T} = 
-\frac{2}{c^2}\int_{a_\mathrm{far}}^{a_\mathrm{near}} 
\frac{\mathrm{d}\Phi}{\mathrm{d}a}\,\mathrm{d}a, 
\end{equation}   
where $a_\mathrm{far}$ is the scale factor when the photon crossed the far
side of the void and $a_\mathrm{near}$ is the scale factor when the
photon crossed the near side of the void. The gravitational potential 
of a void with proper radius $r$ is
\begin{equation}
\Phi=\frac{4\pi G}{3}r^2\rho_\mathrm{b}\delta\left(a\right).
\end{equation}
Here the background density is given by
$\rho_\mathrm{b}=3H^2_0\Omega_\mathrm{m}/8\pi G a^3$ and the density
perturbation is given by
$\delta\left(a\right)=D\left(a\right)\delta\left(a_0\right)$ where $D$
is the linear growth factor. Hence
\begin{equation}
\frac{\Delta T}{T} = 
\Omega_\mathrm{m}\left(\frac{R}{c/H_0}\right)^2 
\left[\frac{D(a_\mathrm{far})}{a_\mathrm{far}} - 
\frac{D(a_\mathrm{near})}{a_\mathrm{near}}\right]\delta. 
\label{isw}
\end{equation}
Using this we calculate the expected ISW signal for the 50 highest
significance voids in Table 4 of \citet{Granett:2008xb}, employing the
concordance $\Lambda$CDM cosmology to determine $a_\mathrm{far}$ and
$a_\mathrm{near}$ for each void from the void redshift
measurements. The ISW signal is found to be only $-0.42 \,\mu
\mathrm{K}$ on average if the dark matter underdensities are smaller
than the observed underdensities in the LRG counts by the bias factor
of 2.2 (taking $\sigma_8=0.8$). This is in contrast to the detected
mean signal of $-11.3 \,\mu \mathrm{K}$ which is over 20 times bigger!
We must therefore conclude that the void radii and/or underdenities
have been significantly underestimated. The void radii can at most be
increased by a factor of 1.75 within the quoted uncertainties so the
observed signal of $-11.3 \,\mu \mathrm{K}$ can be matched only if the
underdensities are increased by a factor of 5 (implying a bias factor
of 0.2). The CMB temperature decrements of such model voids calculated
using eq.(\ref{isw}) are shown in Fig.\ref{hist} and are (by
construction) similar to the actual measurements shown in Fig.2 of
\citet{Granett:2008ju}. While such an {\em underbias} for the observed
LRGs may seem implausible, we emphasise that this is the only way in
which the temperature decrements observed by
\citet{Granett:2008xb,Granett:2008ju} can be accounted for as being
due to the late ISW effect.

Fig.\ref{hist} displays a histogram of the probabilities for finding
such voids in the SDSS LRG survey volume
($5\,h^{-3}\,\mathrm{Gpc}^{3}$ in the redshift range $0.4 < z <
0.75$). The most improbable void is at $z=0.672$ --- in order to yield
the {\em observed} average CMB temperature decrement it must have a density
contrast of -0.72 (quoted galaxy underdensity of -0.316 multiplied by
5/2.2) and a radius of $230~h^{-1}\,\mathrm{Mpc}$ (radius derived from
the quoted volume of $10^7~h^{-3}\,\mathrm{Mpc}^3$ and multiplied by
1.75). The probability of such a void is $1.9 \times 10^{-247}$
according to our calculations. Although linear theory may not be
applicable for such a deep void, it is clear that its existence is in
gross conflict with the standard theory of structure formation from
{\em gaussian} primordial density perturbations.

This conclusion is strengthened by the recent detection of very large
peculiar velocities on large scales. As seen in Fig.\ref{var}, the
expected variance of the peculiar velocity as calculated by
eq.(\ref{var3}) is about 200 km~s$^{-1}$ on a scale of 100 $h^{-1}$
Mpc, whereas the measured value is at least 5 times higher, and the
discrepancy is even bigger on larger scales up to 300 Mpc
\citep{Kashlinsky:2008ut}.

It is also seen from Fig.\ref{prob1} that if a determination of the
Hubble constant is required with say 1\% accuracy, then measurements
extending out to at least $150~h^{-1}\,\mathrm{Mpc}$ must be made to
overcome local fluctuations. A similar estimate was made by
\citet{Li:2008yj} who noted that the observed variance in measurements
of $h$ is in accord, thus consistent with the assumption of a gaussian
density field.  However the voids observed in the SDSS LRG survey
\citep{Granett:2008ju} call this assumption into question. In
particular whether there is a large local void is then an issue that
must be addressed observationally and not dismissed on the grounds
that it is inconsistent with gaussian perturbations. The Hubble flow
is presently poorly measured in the redshift range $0.1 \la z \la 0.3$
--- just where the effects of such a local void would be most apparent
\citep{Alexander:2007xx}. Given that dark energy may well be an
artifact of such a void, this issue needs urgent attention.

The question of how such voids can have been generated without
conflicting with the CMB observations is beyond the scope of the
present work. Some suggestions have been made in the
context of multi-field inflationary models
\citep{Occhionero:1997eh,DiMarco:2005zn,Itzhaki:2008ih}.

\section{acknowledgments}

This work was supported by a STFC Senior Fellowship award to S.S.
(PPA/C506205/1) and by the EU Marie Curie Network ``UniverseNet''
(HPRN-CT-2006-035863).

\end{document}